\documentclass[10pt]{article}
\usepackage{amssymb,amsmath,fullpage,times,epsf}
\usepackage{CJK}
\usepackage{mathrsfs}
\usepackage{stfloats}
\usepackage{lipsum}
\usepackage{amssymb}
\usepackage{subfigure}
\usepackage{float}
\usepackage{graphics}
\usepackage{graphicx}
\usepackage{color}
\usepackage{bm}
\usepackage{amsfonts}
\usepackage{color}
\usepackage{comment}
\usepackage{extarrows}
\usepackage{epsfig}

\DeclareSymbolFont{largesymbol}{OMX}{yhex}{m}{n}
\DeclareMathAccent{\Widehat}{\mathord}{largesymbol}{"62}

\numberwithin{equation}{section}

\allowdisplaybreaks

\newtheorem{theorem}{Theorem}[section]

\newtheorem{lemma}[theorem]{Lemma}

\newtheorem{remark}[theorem]{Remark}

\newcommand{\mi}{\mathrm{i}}

\def\diag{\mathrm{diag}}

\def\sech{\mathrm{sech}}

\def\exp{\mathrm{exp}}
\def\arg{\mathrm{Arg}}
\def\Res{\mathrm{Res}}
\def\det{\mathrm{det}}
\def\max{\mathrm{max}}

\def\eref#1{(\ref{#1})}

\usepackage[square,numbers,sort&compress]{natbib}
\usepackage{caption}
\begin{document}

\date{}

\title{Inverse scattering theory for the discrete PT-symmetric nonlocal nonlinear Schr\"{o}dinger equation under arbitrarily large nonzero boundary conditions}

\author{Chuanxin Xu$^{1}$, Tao Xu$^{1,}$\thanks{Corresponding author, e-mail: xutao@cup.edu.cn.}\\
	{\em 1. College of Science, China University of Petroleum, Beijing
		102249, China} \\}

\maketitle

\date{}
\vspace{-5mm}

\begin{abstract}

In this paper, the theory of inverse scattering transform (IST) is developed for the discrete PT-symmetric nonlocal nonlinear Schr\"{o}dinger equation under large nonzero boundary conditions (NZBCs). By considering that the data at infinity have constant amplitudes, two cases are studied where the previous IST theory fails for large NZBCs. Based on a suitable uniformization variable, the rigorous proofs for the analyticity, symmetries and asymptotic behaviors of the eigenfunctions and scattering coefficients are provided for the direc problem, and the potential reconstruction formula is derived by solving the Riemann-Hilbert problem. Particularly, the focusing equation is found to admit two types of novel solitons under large NZBCs: oscillating soliton and breather, where the former has not been previously reported, while the latter does not occur under small NZBCs. In addition, the multi-soliton solutions are shown to exhibit the collisions among oscillating dark/anti-dark solitons, and the superposition of oscillating soliton and breather.

\vspace{5mm}

	
\end{abstract}

\newpage

\section{Introduction}
In 2013, Ablowitz and Musslimani first introduced the concept of parity-mirror nonlocality into the integrable equation, and proposed the parity-time (PT) symmetric nonlocal nonlinear Schrödinger (NNLS) equation~\cite{Ablowitz1}:
\begin{equation}
	\label{NNLSS1}
	\mi q_t(x,t) = q_{xx}(x,t) - 2 \sigma q^2(x,t) q^*(-x,t),
\end{equation}
where $q(x,t)$ is a complex-valued function of real variables $x$ and $t$, the asterisk represents complex conjugate, and  $\sigma= \pm 1$ correspond to the defocusing $(+)$ and focusing $(-)$ cases, respectively.
It is notable that Eq.~\eref{NNLSS1} is a special case of the Ablowitz-Kaup-Newell-Segur integrable hierarchy under the reduction of PT symmetry, and has the nonlocal nonlinear term depending on $x$ and $-x$.
Over the past ten years, the PT-symmetric NNLS equation has attracted considerable interest, with intensive studies on its mathematical properties~\cite{Ablowitz1,lnverse,Santini,response4,Genoud,feg6,Shepelsky2023,RY1,RY2,RY3,VSA} and exact nonlinear localized-wave solutions~\cite{Ablowitz1,lnverse,YJK3,Ablowitz119,Ablowitz11,FBF,LML, LiXu, Michor, Wen2, YZY, LiXu1,TSF6, Gurses,WPZ,ZDJ}.
On the other hand, a number of new integrable nonlocal nonlinear evolution equations (including 1+1 and 1+2 dimensions and semi-discrete settings) have been discovered by introducing new types of nonlocal reductions, such as the space, time, and space-time reversals and their proper combinations with charge conjugation and space/time translation~\cite{Ablowitz21,Ablowitz119,Ablowitz6,Ablowitz7,Ablowitz8,Fokas,vector,Sinha,DNLS,Rao,mKdV,NNWave,Lou2,Cen,JKY,NKdV,W}.


In 2014, Ablowitz and Musslimani introduced the discretization of Eq.~\eref{NNLSS1}, i.e., the following discrete PT-symmetric nonlocal nonlinear Schr\"{o}dinger (NNLS) equation~\cite{Ablowitz21}:
\begin{equation}
	\label{NNLSS11}
\begin{aligned}
	\mi \frac{d Q_{n}(t) }{d t} = Q_{n+1}(t) -  2 Q_{n}(t) + Q_{n-1}(t) - \sigma Q_{n}(t) Q^*_{-n}(t) \big[ Q_{n+1}(t) + Q_{n-1}(t) \big],
\end{aligned}
\end{equation}
where $Q_{n}(t)$ is a complex-valued function of $n \in \mathbb{Z}$ and real variable $t$. Compared with the celebrated Ablowitz-Ladik equation~\cite{AL1,AL2}, the nonlocal nonlinearity term of Eq.~\eref{NNLSS11} couples the solution values at the lattice sites $n$ and $-n$. Also, Eq.~\eref{NNLSS11} remains invariant under the combined operations of $n \to -n$ and complex conjugation, which means that it satisfies the PT symmetry.
Recently, Eq.~\eref{NNLSS11} has been studied from different mathematical aspects,
like the inverse scattering theory with zero and nonzero backgrounds~\cite{Ablowitz2020, YHLiu,Grahovski},
gauge equivalence between Eq.~\eref{NNLSS11} and the discrete Heisenberg-like equation~\cite{Ma2016b},
discrete nonlocal soliton dynamics on the metric graphs~\cite{Akramov2023}, Darboux transformation (DT) and $N$-soliton solutions in the determinant form~\cite{XuTao2024,LHJ}.
Eq.~\eref{NNLSS11} was solved by some analytical methods, such as the inverse scattering transform (IST)~\cite{Ablowitz2020,YHLiu,Grahovski}, DT~\cite{Ma2016b, Hu2018, Hanif2020, XuTao2024, Yu2025,Hanif2019}, and Hirota bilinear method~\cite{Ma2016a,Qin2025}.

As a nonlinear analogue of the Fourier transform, the IST plays an important role in studying the soliton equations, like the construction of $N$-soliton solutions~\cite{Ablowitz21,Grahovski,Ablowitz2020},
long-time asymptotics~\cite{Fan2}, existence of global solutions~\cite{feg6} and asymptotic stability of $N$-soliton solutions~\cite{Tian3}.
In 2014, Ref.~\cite{Ablowitz21} first established the IST scheme of Eq.~\eref{NNLSS11} with the zero boundary conditions (ZBCs), and obtained the singular one-soliton solution.
Then, Eq.~\cite{Grahovski} derived the two-soliton solution based on the IST formulation with ZBCs.
Futhermore, Ref.~\cite{Ablowitz2020} gave a detailed study for the IST of PT-symmetric, reverse-space-time, and reverse-time discrete NLS equations with ZBCs, and found that their self-similar reductions can yield discrete nonlocal Painlevé-type equations.
Very recently,  Ref.~\cite{YHLiu} developed the IST theory of Eq.~\eref{NNLSS11} with nonzero boundary conditions (NZBCs),
and derived the bright-, dark- and dark-bright soliton solutions.
However, the IST in Ref.~\cite{YHLiu} was studied only for small boundary conditions with the background amplitude less than 1, so that no breather solution was found.
In fact, an earlier study~\cite{Ablowitz2007} on the IST for the Ablowitz-Ladik (AL) equation showed that small boundary conditions confines the discrete eigenvalues $z$ of Lax pair to the circle $|z|=1$.
Particularly for the defocusing AL equation with large NZBCs (i.e., the background amplitude is greater than 1), Ref.~\cite{Alyssa} showed that there is no restriction on the location of discrete eigenvalues, and obtained the discrete Kuznetsov-Ma, Akhmediev and Peregrine solutions.

In this paper, we generalize the IST for Eq.~\eref{NNLSS11} with arbitrarily large NZBCs:
\begin{equation}
	\begin{aligned}
		&\displaystyle\lim_{n \to \pm \infty} Q_{n}(t) = Q_{\pm} e^{2 \mi \sigma Q_{+} Q^*_{-} t }, \quad Q_{\pm} = Q_{0} e^{\mi  \theta_{\pm}} (0 \leq \theta_{\pm} < 2\pi),
	\end{aligned}
\end{equation}
where $Q_0>1$ represents large NZBCs. Similarly to the continuous case~\cite{Ablowitz11}, if $\Delta \theta:= \theta_{+}-\theta_{-} \neq 0$ and $\pi$, the amplitude of $Q_n(t)$ will exponentially grow or decay as $t \to \pm \infty$.
So, the solution remains bounded only when $\Delta \theta = 0$ or $\pi$ is satisfied.
To make the boundary conditions independent of $t$, we perform the transformation $Q_n(t) \to \hat{Q}_n(t) e^{-2 \mi \sigma Q_{+} Q^*_{-} t}$ and write the equation and boundary conditions as follows:
\begin{subequations}
\begin{align}
	&\mi \frac{d Q_{n}(t) }{d t} = Q_{n+1}(t) +  2 r^2 Q_{n}(t) + Q_{n-1}(t) - \sigma Q_{n}(t) \big[   Q^*_{-n}(t) \big( Q_{n+1}(t) + Q_{n-1}(t) \big)   \big] =0 ,\label{NNLS1a} \\
	&\displaystyle\lim_{n \to \pm \infty} Q_{n}(t) = Q_{\pm} = Q_{0} e^{\mi  \theta_{\pm}}, \label{NZBC}
\end{align}
\end{subequations}
where $r^2 : = \sigma Q_{+} Q^*_{-} - 1 \in \mathbb{R}$ and the hat has been dropped for brevity.
For $\sigma e^{\mi \Delta \theta} = -1$, the IST theory developed in Ref.~\cite{YHLiu} can also apply to both small $(0<Q_0<1)$ and large boundary conditions ($Q_0 > 1$).
In this paper, we focus on the case $\sigma e^{\mi \Delta \theta} = 1$ under large boundary conditions  $Q_0^2 = \sigma Q_{+} Q^*_{-} > 1$, which includes two subcases: (i) $\sigma = 1$ and $\Delta \theta = 0$;
(ii) $\sigma = -1$ and $\Delta \theta = \pi$.
Motivated by the work of Ref.~\cite{Alyssa}, we formulate the direct problem using a suitable uniformization variable, and construct the corresponding Riemann-Hilbert problem (RHP) for the inverse problem.
We provide the proofs for the analyticity of the eigenfunctions, symmetries and asymptotic behaviors of the eigenfunctions and scattering coefficients, and derive the potential reconstruction formula with the presence of an arbitrary number of simple zeros.
Moreover, we present the determinant representation of $N$-soliton solutions for arbitrarily large NZBCs. In particular, the solutions for the focusing case of Eq.~\eref{NNLS1a} admit two types of novel solitons: (i) The oscillating soliton, corresponding to an eigenvalue lying on the unit circle $\Gamma: |z| = 1$, can display the oscillatory dark or anti-dark profile along the lattice sites $n$; (ii) The breather, associated with a pair of eigenvalues symmetrically located with respect to $\Gamma$,  features internal periodic oscillation as the time $t$ evolves.
It is worth noting that the oscillating soliton has not been previously reported for either the PT-symmetric NNLS equation~\cite{Ablowitz11} or the AL equation~\cite{,Ablowitz2007,Alyssa}, while the breather does not occur in the case of small NZBCs~\cite{YHLiu}.


The rest of this paper is organized as follows:
Section~\ref{sec2} studies the direct scattering problem, including the analyticity and symmetries of the eigenfunctions and scattering coefficients.
Based on the RHP, section~\ref{sec3} addresses the pole contributions, constraints of norming constants, trace formula, and reconstruction formula.
In section~\ref{sec4}, we present the determinant representation of $N$-soliton solutions, and discuss the dynamical properties of solutions.
Finally, section~\ref{sec5} provides the conclusions of this paper.

%


\section{Direct scattering problem}
\label{sec2}
This section studies the direct scattering problem of Eq.~\eref{NNLS1a} under large NZBCs~\eref{NZBC}, including the introduction of the uniformization variable and the analysis of analyticity, symmetries and asymptotic behavior of the eigenfunctions and scattering coefficients.

\subsection{Lax pair}
\label{Sec1}
Eq.~\eref{NNLS1a} has the following Lax pair:
\begin{subequations}
	\label{hlj9a}
	\begin{align}
		& \bm{\phi}_{n+1}(t,z)
		= \mathbf{U}_{n}(t,z) \bm{\phi}_{n}(t,z),   \label{Laxpairjvzhenxingshiaa}   \\
		& \frac{d\bm{\phi}_{n}(t,z) }{d t}
		= \mathbf{V}_{n}(t,z) \bm{\phi}_{n}(t,z), \label{Laxpairjvzhenxingshibb}
	\end{align}
\end{subequations}
with
\begin{subequations}
	\begin{align}
		&\mathbf{U}_{n}(t,z) = \left(
		\begin{array}{ccc}
			z & Q_{n}(t) \\
			\sigma Q^*_{-n}(t) & \frac{1}{z} \\
		\end{array}
		\right),     \\
		&\mathbf{V}_{n}(t,z) = \left(
		\begin{array}{ccc}
			\mi \sigma Q_{n}(t) Q^*_{-n+1}(t)  - \mi (r^2 + z^2) & - \mi z Q_{n}(t) +  \frac{\mi}{z} Q_{n-1}(t) \\
			 \frac{\mi}{z} \sigma Q^*_{-n}(t)- \mi z \sigma Q^*_{-n+1}(t)   & - \mi \sigma Q_{n-1}(t) Q^*_{-n}(t)  + \mi (r^2 +\frac{1}{z^2} )
		\end{array}
		\right),
	\end{align}
\end{subequations}
where $\bm{\phi}_{n}(t,z)$ is the $2 \times 2$ matrix solution, $z \in \mathbb{C}$ is the spectral parameter. Eq.~\eref{NNLS1a} is equivalent to the zero-curvature condition $\frac{d \mathbf{U}_{n}(t,z)}{d t} = \mathbf{V}_{n+1}(t,z) \mathbf{U}_{n}(t,z) - \mathbf{U}_{n}(t,z) \mathbf{V}_{n}(t,z)$ (also known as the compatibility condition $\frac{d}{dt} \bm{\phi}_{n+1}(t,z) = \frac{d}{dt} \bm{\phi}_{n}(t,z)|_{n\to n+1}$).


Taking the limits of Eqs.~\eref{hlj9a} as $n\to \pm \infty$ yields the following asymptotic scattering problem:
\begin{equation}
	\label{sg5a}
	\begin{aligned}
		&\bm{\phi}_{n+1}(t,z) = \mathbf{U}_{\pm}(z) \bm{\phi}_{n}(t,z), \quad
		\frac{d\bm{\phi}_{n}(t,z) }{d t} = \mathbf{V}_{\pm}(z) \bm{\phi}_{n}(t,z),                                     \\
	\end{aligned}
\end{equation}
where
\begin{equation}
	\label{xcxa10a}
	\begin{aligned}
		& \mathbf{U}_{\pm}(z) = \left(
		\begin{array}{ccc}
			z & Q_{\pm}  \\
			\sigma Q^*_{\mp} & \frac{1}{z} \\
		\end{array}
		\right),  \,\,\\
		&\mathbf{V}_{\pm}(z) = \left(
		\begin{array}{ccc}
			\mi  - \mi z^2  & - \mi Q_{\pm} \big(z - \frac{1}{z} \big) \\
			- \mi \sigma Q^{*}_{\mp} \big(z - \frac{1}{z} \big)   & - \mi  + \frac{\mi}{z^2}  \\
		\end{array}
		\right).
	\end{aligned}
\end{equation}
Since the matrices $\mathbf{U}_{\pm}(z)$ and  $\mathbf{V}_{\pm}(z)$ are both independent of $n$ and $t$, the compatibility condition of Eqs.~\eref{sg5a} becomes $[\mathbf{U}_{\pm}(z), \mathbf{V}_{\pm}(z)] = 0$, which means that $\mathbf{U}_{\pm}(z)$ and $\mathbf{V}_{\pm}(z)$ have the same eigenvectors. The eigenvalues of $\mathbf{U}_{\pm}(z)$ and $\mathbf{V}_{\pm}(z)$ are given by
\begin{subequations}
	\label{hlj3.9}
	\begin{align}
		& \mathbf{U}_{\pm}(z) : \, \Bigl\{ \frac{1}{2} \big(z + \frac{1}{z}  + \frac{k(z)}{z} \big), \, \frac{1}{2} \big(z + \frac{1}{z} - \frac{k (z)}{z} \big)\Bigr\}, \label{hlj3.9a}\\
		& \mathbf{V}_{\pm}(z) : \, \Bigl\{-\frac{\mi}{2} \big(z - \frac{1}{z} \big) \big(z + \frac{1}{z}  + \frac{k (z) }{z} \big), \, -\frac{\mi}{2} \big(z - \frac{1}{z} \big) \big(z + \frac{1}{z} - \frac{k (z) }{z} \big)\Bigr\},
	\end{align}
\end{subequations}
with $k (z)  = \sqrt{(z^2-1)^2 + 4 z^2  Q^2_0} $, and their common eigenvector matrix is
\begin{equation}
	\label{hlj4.0}
	\begin{aligned}
		\mathbf{Y}_{\pm}(z)
		=\left(
		\begin{array}{ccc}
			Q_{\pm}  & \frac{1 - z^2 + k(z) }{2 z }    \\
			\frac{1 - z^2 + k(z) }{2 z }   &  - \sigma Q^*_{\mp} \\
		\end{array}
		\right).
	\end{aligned}
\end{equation}

\begin{figure}[H]
 	\centering
 		\label{f0a}
 		\includegraphics[width=2.9in]{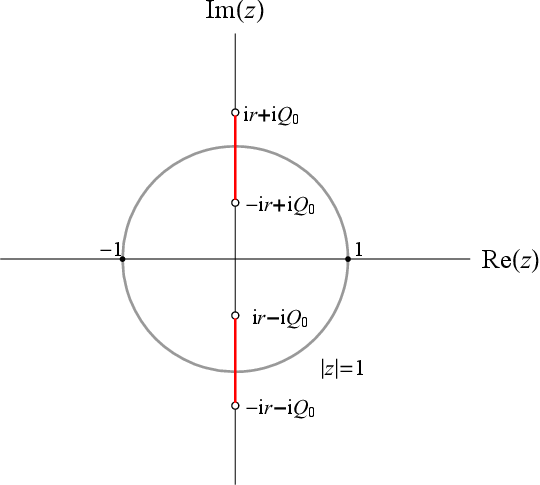}\hfill
 	\caption{\small
 		The complex $z$-plane, including four branch points $z = \mi r  \pm \mi Q_0, - \mi r  \pm \mi Q_0$ and branch cut $(- \mi r  - \mi Q_0,  \mi r  - \mi Q_0) \cup (- \mi r  + \mi Q_0,  \mi r  + \mi Q_0)$.  \label{f01} }
 \end{figure}

To eliminate the square root operation of $k(z)$, we introduce the following transformation
\begin{equation}
	\label{hlj17}
	\begin{aligned}
		z + \frac{1}{z}= \mi r \Big( \lambda +\frac{1}{\lambda} \Big),
	\end{aligned}
\end{equation}
so that $\lambda(z)$ can be inversely expressed as
\begin{equation}
	\label{hlj3.11}
	\begin{aligned}
		\lambda(z) = \xi(z) \pm \sqrt{\xi(z)^2 - 1},
	\end{aligned}
\end{equation}
with $\xi(z) = \frac{1}{2 \mi r} \big(z +\frac{1}{z} \big)$.
Here, it is obvious that $\lambda(z)$ has the branching and the branch points  are located at $\xi^2 = 1$, i.e.,
\begin{equation}
	\label{hlj3.12}
	\begin{aligned}
		\Big(z +\frac{1}{z} \Big)^2 = 4 (\mi r)^2 \mapsto z^2 \mp 2 \mi r z + 1 = 0,
	\end{aligned}
\end{equation}
which implies that the four branch points are given by $z = \mi r  \pm \mi Q_0, - \mi r  \pm \mi Q_0$.
Meanwhile, there must be $|\lambda(\xi)| = 1$ if $\xi \in [-1,1]$.
Accordingly, we can define the Riemann surface of the equation $(\lambda-\zeta(z))^2 = \zeta^2(z) - 1$, which consists of the first sheet $\lambda = \xi(z) + \sqrt{\xi(z)^2 - 1}$ and the second sheet $\lambda  = \xi(z) - \sqrt{\xi(z)^2 - 1}$.
As shown in Fig.~\ref{f01}, $\lambda$ has four branch points at $z = \mi r  \pm \mi Q_0, - \mi r  \pm \mi Q_0$, and there exists the branch cut $z \in (- \mi r  - \mi Q_0,  \mi r  - \mi Q_0) \cup (- \mi r  + \mi Q_0,  \mi r  + \mi Q_0)$ in the $z$-plane.

As a result, the eigenvalues of $\mathbf{U}_{\pm}$ and $\mathbf{V}_{\pm}$ can be expressed as
\begin{equation}
	\label{hlj4.1}
	\begin{aligned}
		\mathbf{U}_{\pm}(z,\lambda): \,\Bigl\{\mi r \lambda, \, \frac{\mi r}{\lambda}\Bigr\}; \quad
		\mathbf{V}_{\pm}(z,\lambda): \, \Bigl\{r \lambda  \Big(z - \frac{1}{z}\Big), \, \frac{r}{\lambda} \Big(z - \frac{1}{z}\Big)\Bigr\},
	\end{aligned}
\end{equation}
and the eigenvector matrix becomes
\begin{equation}
	\begin{aligned}
		\mathbf{Y}_{\pm}(z,\lambda)
	    =\left(
		\begin{array}{ccc}
			Q_{\pm}  &  \mi r \lambda - z   \\
			\mi r \lambda - z   &  -\sigma Q^*_{\mp} \\
		\end{array}
		\right),
	\end{aligned}
\end{equation}
whose determinant is $\gamma(z,\lambda): = - Q^2_0 - (\mi r \lambda - z)^2 $.

\subsection{Eigenfunctions and their analyticity in the $\lambda$-plane}
As the simultaneous solutions of the Lax pair~\eref{hlj9a}, $\bm{\phi}_{n}^{\pm}(t,z,\lambda)$ satisfy the asymptotic behavior:
\begin{equation}
	\label{hlj21}
	\begin{aligned}
		& \bm{\phi}_{n}^{\pm}(t,z,\lambda) \sim \mathbf{Y}_{\pm}(z,\lambda) (\mi r \mathbf{J}(z,\lambda))^n e^{\mathbf{\Omega}(z,\lambda) t}, \,\, n \to \pm \infty,
	\end{aligned}
\end{equation}
with $\mathbf{J}(z,\lambda)  = \diag\big( \lambda, \frac{1}{\lambda} \big)$ and $\mathbf{\Omega}(z,\lambda)  = (\mathrm{\Omega}_{ij}(z,\lambda)) = \diag \Big(r\lambda \big(z - \frac{1}{z} \big), \frac{r}{\lambda} \big(z - \frac{1}{z} \big)  \Big)$.
To normalize the asymptotic behavior of $\bm{\phi}_{n}^{\pm}(t,z,\lambda)$, we introduce the following modified eigenfunctions
\begin{equation}
	\label{xcxa19}
	\begin{aligned}
		&\bm{\mu}_{n}^{\pm}(t,z,\lambda) = \mathbf{Y}_{\pm}^{-1}(z,\lambda)  \bm{\phi}_{n}^{\pm}(t,z,\lambda)
		(\mi r \mathbf{J}(z,\lambda))^{-n} e^{-\mathbf{\Omega}(z,\lambda) t},
	\end{aligned}
\end{equation}
which satisfy the asymptotic behavior $\displaystyle\lim_{n \to \pm \infty}\bm{\mu}_{n}^{\pm}(t,z,\lambda) = \mathbf{I}$.
Substituting Eq.~\eref{xcxa19} into Eq.~\eref{Laxpairjvzhenxingshiaa} yields
\begin{subequations}
	\label{xcxa20}
	\begin{align}
		&\bm{\mu}_{n+1}^{-}(t,z,\lambda)
		\mi r \mathbf{J}(z,\lambda)
		=  \big(
		\mi r \mathbf{J}(z,\lambda)
		+
		\mathbf{Y}^{-1}_{-}(z,\lambda) \Delta \mathbf{Q}_{n}^{-}(t)  \mathbf{Y}_{-}(z,\lambda) \big) \bm{\mu}_{n}^{-}(t,z,\lambda), \label{xcxa20aa} \\
		&\bm{\mu}_{n+1}^{+} (t,z,\lambda)
		\mi r \mathbf{J}(z,\lambda)
		=  \big(
		\mi r \mathbf{J}(z,\lambda)
		+
		\mathbf{Y}^{-1}_{+}(z,\lambda) \Delta \mathbf{Q}_{n}^{+}(t)  \mathbf{Y}_{+}(z,\lambda) \big) \bm{\mu}_{n}^{+}(t,z,\lambda), \label{xcxa20ab}
	\end{align}
\end{subequations}
where $\Delta \mathbf{Q}_{n}^{\pm}(t) = \mathbf{Q}_{n}(t) - \mathbf{Q}_{n}^{\pm}$ with $\mathbf{Q}_{n}(t) = \diag(Q_{n}, \sigma Q^*_{n})$ and $\mathbf{Q}_{n}^{\pm} = \diag(Q_{\pm}, \sigma Q^*_{\mp})$.
Here, since the asymptotic behavior of $\bm{\mu}_{n}^{+} (t,z,\lambda)$ is known only as $n \to +\infty$, the function $\bm{\mu}_{n}^{+}(t,z,\lambda)$ cannot be determined from Eq.~\eref{xcxa20ab}.
To address this issue, we write this equation in the backward form (see the proof in Appendix~\ref{appendixA1a}):
\begin{equation}
	\label{xcxa21}
	\begin{aligned}
		&\hat{\bm{\mu}}_{n}^{+}(t,z,\lambda)
		(\mi r \mathbf{J}(z,\lambda))^{-1} =
		\big(
		(\mi r \mathbf{J}(z,\lambda))^{-1}
		-   \mathbf{Y}^{-1}_{+}(z,\lambda)  \Delta \mathbf{Q}_{n}^{+}(t) \mathbf{Y}_{+}(z,\lambda)  \big)  \hat{\bm{\mu}}_{n+1}^{+}(t,z,\lambda),
	\end{aligned}
\end{equation}
with $\hat{\bm{\mu}}_{n}^{+}(t,z,\lambda)  = \bm{\mu}_{n}^{+}(t,z,\lambda) \displaystyle\prod_{k = n}^{ \infty} (1 - \sigma Q_k(t)  Q^*_{-k}(t))$, which implies that $\hat{\bm{\mu}}_{n}^{+}(t,z,\lambda)$ shares the same asymptotic behaviors and analyticity as $\bm{\mu}_{n}^{+}(t,z,\lambda)$.
Then, Eqs.~\eref{xcxa20aa} and~\eref{xcxa21} can be equivalently written in the summation form, which are the discrete versions of the integral equations:
\begin{subequations}
	\label{xcx2.17}
	\begin{align}
		&\mu_{n,1}^{-}(t,z,\lambda) = \left(\begin{array}{ccc}
			1 \\
			0 \\
		\end{array} \right)+
		\displaystyle\sum_{k = -\infty}^{n-1}
		\frac{1}{\mi r\lambda}
		\left( \begin{array}{ccc}
			1 & 0\\
			0 & \lambda^{-2(n-k-1)}\\
		\end{array} \right)
		\mathbf{Y}^{-1}_{-}(z,\lambda) \Delta \mathbf{Q}_{k}^{-}(t) \mathbf{Y}_{-}(z,\lambda) \mu_{k,1}^{-}(t,z,\lambda), \label{xcxa20a}\\
		&\mu_{n,2}^{-}(t,z,\lambda)= \left(\begin{array}{ccc}
			0 \\
			1 \\
		\end{array} \right)+
		\displaystyle\sum_{k = -\infty}^{n-1}
		\frac{\lambda}{\mi r}
		\left( \begin{array}{ccc}
			\lambda^{2(n-k-1)} & 0\\
			0 & 1\\
		\end{array} \right)
		\mathbf{Y}^{-1}_{-}(z,\lambda) \Delta \mathbf{Q}_{k}^{-}(t) \mathbf{Y}_{-}(z,\lambda) \mu_{k,2}^{-}(t,z,\lambda), \label{xcxa20c}
	\end{align}
\end{subequations}
and
\begin{subequations}
	\label{xcx2.18}
	\begin{align}
		&\hat{\mu}_{n,1}^{+}(t,z,\lambda) = \left(\begin{array}{ccc}
			1 \\
			0 \\
		\end{array} \right)-
		\displaystyle\sum_{k = n}^{\infty}
		\mi r\lambda
		\left( \begin{array}{ccc}
			1 & 0\\
			0 & \lambda^{-2(n-k)}\\
		\end{array} \right)
		\mathbf{Y}^{-1}_{+}(z,\lambda) \Delta \mathbf{Q}_{k}^{+}(t) \mathbf{Y}_{+}(z,\lambda)  \hat{\mu}_{k,1}^{+}(t,z,\lambda),\\
		&\hat{\mu}_{n,2}^{+}(t,z,\lambda) = \left(\begin{array}{ccc}
			0 \\
			1 \\
		\end{array} \right)-
		\displaystyle\sum_{k = n}^{\infty}
		\frac{\mi r}{\lambda}
		\left( \begin{array}{ccc}
			\lambda^{2(n-k)} & 0\\
			0 & 1\\
		\end{array} \right)
		\mathbf{Y}^{-1}_{+}(z,\lambda) \Delta \mathbf{Q}_{k}^{+}(t) \mathbf{Y}_{+}(z,\lambda)  \hat{\mu}_{k,2}^{+}(t,z,\lambda),
	\end{align}
\end{subequations}
where $\mu_{n,i}^{-}(t,z,\lambda)$ and $\hat{\mu}_{n,i}^{+}(t,z,\lambda)$ ($i=1,2$) represent the $i$th columns of $\bm{\mu}_{n}^{-}(t,z,\lambda)$ and $\hat{\bm{\mu}}_{n}^{+}(t,z,\lambda)$, respectively.
Based on Eqs.~\eref{xcx2.17} and~\eref{xcx2.18}, we can obtain the analyticity of the columns of $\bm{\mu}_{n}^{\pm}(t,z,\lambda)$ (see the proof in Appendix~\ref{appendixA3}) as follows:
\begin{theorem}
	\label{th1}
	Under the condition $\Delta \mathbf{Q}_{n}^{\pm}(t) \in \Big\{ f_n \big| \displaystyle\sum_{j = \mp \infty}^{n}  | f_{j} | < \infty, \forall n \in \mathbb{Z} \Big\}$, the eigenfunctions $\mu^{-}_{n,1}(t,z,\lambda)$ and $\mu^{+}_{n,2}(t,z,\lambda)$ are  analytic for $|\lambda|>1$, while the eigenfunctions $\mu^{-}_{n,2}(t,z,\lambda)$ and $\mu^{+}_{n,1}(t,z,\lambda)$ are analytic for $|\lambda|<1$. Moreover, both $\mu^{\pm}_{n,1}(t,z,\lambda)$ and $\mu^{\pm}_{n,2}(t,z,\lambda)$ can be extended continuously to the circle $|\lambda|=1$.
\end{theorem}

Taking the determinants for both sides of Eq.~\eref{Laxpairjvzhenxingshiaa} shows the following relation
\begin{equation}
	\begin{aligned}
		\det(\bm{\phi}_{n+1}(t,z,\lambda)) = \big(1 -  \sigma Q_{n}(t) Q^*_{-n}(t)  \big) \det(\bm{\phi}_{n}(t,z,\lambda)).
	\end{aligned}
\end{equation}
Then, considering the asymptotic behavior of $\bm{\phi}^{\pm}_{n}(t,z,\lambda)$ in Eq.~\eref{hlj21}, we have
\begin{equation}
	\label{hlj20}
	\begin{aligned}
		&\big{|}	(\mi r)^{-n}  e^{-\Omega_{11}(z,\lambda) t} \phi^{\pm}_{n,1}(t,z,\lambda), (\mi r)^{-n} e^{-\Omega_{22}(z,\lambda) t} \phi^{\pm}_{n,2}(t,z,\lambda)  \big{|}
		\to  \gamma(z,\lambda), \quad n \to \pm \infty,
	\end{aligned}
\end{equation}
which leads to the determinants of $\phi_n^{\pm}$ as follows:
\begin{equation}
	\label{hlj24}
	\begin{aligned}
		&\det(\phi^{-}_{n}(t,z,\lambda) ) = \gamma(z,\lambda) (\mi r)^{2n} e^{[\Omega_{11}(z,\lambda)+\Omega_{22}(z,\lambda)] t }  \displaystyle\prod_{k = - \infty}^{n-1}
		\frac{\sigma Q^*_{-k}(t) Q_{k}(t) -1  }{Q^2_{0} - 1}, \\
		&\det(\phi^{+}_{n}(t,z,\lambda) ) = \gamma(z,\lambda) (\mi r)^{2n} e^{[\Omega_{11}(z,\lambda)+\Omega_{22}(z,\lambda)] t } \displaystyle\prod_{k = n}^{+ \infty}
		\frac{Q^2_{0} - 1 }{\sigma Q^*_{-k}(t) Q_{k}(t) - 1}. \\
	\end{aligned}
\end{equation}

Meanwhile, because $\bm{\phi}_{n}^{+}(t,z,\lambda)$ and $\bm{\phi}_{n}^{-}(t,z,\lambda)$ are both the fundamental solutions of Lax pair~\eref{hlj9a} for $z\in \mathbb{C}$ with the exception of the branch points $z = \mi r  \pm \mi Q_0, - \mi r  \pm \mi Q_0$, there exists an invertible matrix $\mathbf{S}(z,\lambda)$ such that
\begin{equation}
	\begin{aligned}
		\label{hlj26}
		&\bm{\phi}_{n}^{-}(t,z,\lambda) = \bm{\phi}_{n}^{+}(t,z,\lambda)\mathbf{S}(z,\lambda),
	\end{aligned}
\end{equation}
where $\mathbf{S}(z,\lambda)= (s_{ij}(z,\lambda))_{1\leq i,j \leq 2}$ is called the scattering matrix.
By writing Eq.~\eref{hlj26} in the column-wise form
\begin{equation}
	\label{hlj2.23}
	\begin{aligned}
		&\phi^{-}_{n,1}(t,z,\lambda) = s_{11}(z,\lambda) \phi^{+}_{n,1}(t,z,\lambda) +  s_{21}(z,\lambda) \phi^{+}_{n,2}(t,z,\lambda),  \\
		&\phi^{-}_{n,2}(t,z,\lambda) = s_{12}(z,\lambda) \phi^{+}_{n,1}(t,z,\lambda) +  s_{22}(z,\lambda) \phi^{+}_{n,2}(t,z,\lambda),
	\end{aligned}
\end{equation}
where $\phi_{n,i}^{\pm}(t,z,\lambda)$ represent the $i$th columns of $\bm{\phi}_{n}^{\pm}(t,z,\lambda)$, the elements of $\mathbf{S}(z,\lambda)$ can be solved as
\begin{equation}
	\label{hlj28}
	\begin{aligned}
		&s_{11}(z,\lambda) = \frac{\big{|}
			\phi^{-}_{n,1}(t,z,\lambda), \phi^{+}_{n,2}(t,z,\lambda)\big{|}
		}
		{\big{|}\phi^{+}_{n,1}(t,z,\lambda), \phi^{+}_{n,2}(t,z,\lambda)\big{|}},  \quad
		s_{21}(z,\lambda) = \frac{\big{|}\phi^{+}_{n,1}(t,z,\lambda), \phi^{-}_{n,1}(t,z,\lambda)\big{|}}
		{\big{|}\phi^{+}_{n,1}(t,z,\lambda), \phi^{+}_{n,2}(t,z,\lambda)\big{|}}, \\
		&s_{12}(z,\lambda) = \frac{\big{|}\phi^{-}_{n,2}(t,z,\lambda), \phi^{+}_{n,2}(t,z,\lambda)\big{|}}
		{\big{|}\phi^{+}_{n,1}(t,z,\lambda), \phi^{+}_{n,2}(t,z,\lambda)\big{|}},  \quad
		s_{22}(z,\lambda) = \frac{\big{|}\phi^{+}_{n,1}(t,z,\lambda), \phi^{-}_{n,2}(t,z,\lambda)\big{|}}
		{\big{|}\phi^{+}_{n,1}(t,z,\lambda), \phi^{+}_{n,2}(t,z,\lambda)\big{|}}.
	\end{aligned}
\end{equation}
Further, by using Eq.~\eref{hlj24}, we can obtain
\begin{equation}
	\label{hlj2.25}
	\begin{aligned}
		\det(\mathbf{S}(z,\lambda)) = \displaystyle\prod_{k = - \infty}^{ \infty}
		\frac{\sigma Q^*_{-k}(t) Q_{k}(t) - 1 }{Q^2_{0}  - 1 } = : c_{\infty},
	\end{aligned}
\end{equation}
and rewrite $s_{11}(z,\lambda)$ and $s_{22}(z,\lambda)$ as
\begin{equation}
	\begin{aligned}
		&s_{11}(z,\lambda)
		=\frac{\big{|}Y_{-}(z,\lambda)  \mu^{-}_{n,1}(t,z,\lambda), Y_{+}(z,\lambda) \mu^{+}_{n,2}(t,z,\lambda) \big{|}}
		{\big{|}Y_{+}(z,\lambda) \mu^{+}_{n,1}(t,z,\lambda), Y_{+}(z,\lambda) \mu^{+}_{n,2}(t,z,\lambda)\big{|}}, \\
		&s_{22}(z,\lambda) = \frac{\big{|} Y_{+}(z,\lambda) \mu^{+}_{n,1}(t,z,\lambda) , Y_{-}(z,\lambda)  \mu^{-}_{n,2}(t,z,\lambda)  \big{|}}
		{ \big{|} Y_{+}(z,\lambda) \mu^{+}_{n,1}(t,z,\lambda) ,  Y_{+}(z,\lambda)  \mu^{+}_{n,2}(t,z,\lambda) \big{|}},
	\end{aligned}
\end{equation}
which imply that $s_{11}(z,\lambda)$ and $s_{22}(z,\lambda)$ are analytic for $|\lambda|>1$ and $|\lambda|<1$, respectively.


\subsection{Uniformization variable and associated eigenfunctions}
Noticing from Eq.~\eref{hlj3.11} the eigenfunctions and scattering data are multivalued functions of $z$, being defined on the two-sheeted Riemann surface for $\lambda(z)$, we introduce the following uniformization variable~\cite{Ablowitz2007, Alyssa}:


\begin{equation}
	\label{hlj29}
	\begin{aligned}
		\zeta(z,\lambda)  = \frac{\lambda}{z}.
	\end{aligned}
\end{equation}
Correspondingly, there are the following inverse mappings:
\begin{equation}
	\label{hlj30}
	\begin{aligned}
		&z^2 = \frac{\zeta - \mi r}{\zeta (\mi r \zeta -1)}, \quad
		\lambda^2 =   \frac{\zeta(\zeta - \mi r)}{\mi r \zeta -1}, \quad
		z \lambda = \frac{\zeta - \mi r}{\mi r \zeta -1}, \\
	\end{aligned}
\end{equation}
which mean that all even powers of $\lambda$ and $z$  can be expressed in terms of $\zeta$.
Meanwhile, Eq.~\eref{hlj30} implies the following relations:
\begin{subequations}
	\begin{align}
		&|\lambda|^2 = 1 \Rightarrow  \big(|\zeta|^2 + 1\big)\big(|\zeta|^2 - 1 + \mi r(\zeta-\zeta^* ) \big) = 0 \Rightarrow |\zeta - \mi r|^2 = Q^2_0, \\
		&|z|^2 = 1 \Rightarrow        \big(|\zeta|^2 + 1\big)\big(r^2 |\zeta|^2 - r^2 - \mi r(\zeta-\zeta^* ) \big) = 0 \Rightarrow |\mi r \zeta + 1|^2 = Q^2_0.
	\end{align}
\end{subequations}
Thus, we know that the set of the continuous spectrum $|\lambda| = 1$ can be mapped onto $\Sigma = \{\zeta \in \mathbb{C}: |\zeta - \mi r|^2 = Q^2_0\}$.
For convenience, we define the following regions:
\begin{equation}
	\begin{aligned}
		&D^{+} = \{\zeta \in \mathbb{C}:  |\zeta - \mi r|^2 < Q^2_0 \}, \quad D^{-} = \{\zeta \in \mathbb{C}:  |\zeta - \mi r|^2 > Q^2_0 \},
	\end{aligned}
\end{equation}
which correspond to $|\lambda|<1$ and $|\lambda|>1$, respectively.
Meanwhile, we show the mappings for some special points from the complex $z$-plane to the $\xi$/$\lambda$/$\zeta$-planes with sheet $\mathrm{I}$: $\lambda = \xi + \sqrt{\xi^2 - 1}$ (see Table~\ref{Table1}) and with sheet $\mathrm{II}$: $\lambda = \xi - \sqrt{\xi^2 - 1}$ (see Table~\ref{Table2}).
\begin{table}[h]
	\small
	\caption{Mapping of some particular points from the $z$-plane to the $\xi$-, $\lambda$- and $\zeta$-planes for Sheet $\mathrm{I}$.
		 \label{Table1}}
	\begin{center}
		\begin{tabular}{|c|c|c|c|c|c|}
			\hline
			$z$   &   $\xi$ & $\lambda$ & $\zeta$
			\\  \hline
			$\mi$   &  $0$ &   $\mi $  &  $1$
			\\  \hline
			$-\mi$   &  $0$ &   $\mi $  &  $-1$
			\\  \hline
			$1$   &  $-\frac{\mi}{r}$ &   $\frac{\mi}{r} \big(Q_0-1\big)$ &  $\frac{\mi }{r} \big(Q_0-1\big)$
			\\  \hline
			$-1$   &  $\frac{\mi}{r}$ &   $\frac{\mi}{r} \big(Q_0+1\big)$  &  $\frac{\mi }{r} \big(Q_0+1\big)$
			\\  \hline
			$\mi r+\mi Q_0$   &  1      &   1&     $-\frac{\mi}{r+Q_0}$
			\\  \hline
			$\mi r-\mi Q_0$   &  1      &   1&     $-\frac{\mi}{r-Q_0}$
			\\  \hline
			$-\mi r+\mi Q_0$   &  $-1$      &  $-1$&    $-\frac{\mi}{r-Q_0}$
			\\  \hline
			$-\mi r-\mi Q_0$   &  $-1$      &   $-1$&    $-\frac{\mi}{r+Q_0}$
			\\  \hline
		\end{tabular}
	\end{center}
\end{table}
\begin{table}[h]
	\small
	\caption{Mapping of some particular points from the $z$-plane to the $\xi$-, $\lambda$- and $\zeta$-planes for Sheet $\mathrm{II}$. \label{Table2}}
	\begin{center}
		\begin{tabular}{|c|c|c|c|c|c|}
			\hline
			$z$   &   $\xi$ & $\lambda$ & $\zeta$
			\\  \hline
			$\mi$   &  $0$ &   $-\mi$  &  $-1$
			\\  \hline
			$-\mi$   &  $0$ &   $-\mi$  &  $1$
			\\  \hline
			$1$   &  $-\frac{\mi}{r}$ &   $-\frac{\mi}{r} \big(Q_0+1\big)$ &  $-\frac{\mi}{r} \big(Q_0+1\big)$
			\\  \hline
			$-1$   &  $\frac{\mi}{r}$ &   $-\frac{\mi}{r} \big(Q_0-1\big)$  &  $-\frac{\mi}{r} \big(Q_0-1\big)$
			\\  \hline
			$\mi r+\mi Q_0$   &  1      &   1&     $-\frac{\mi }{r+Q_0}$
			\\  \hline
			$\mi r-\mi Q_0$   &  1      &   1&     $-\frac{\mi }{r-Q_0}$
			\\  \hline
			$-\mi r+\mi Q_0$   &  $-1$      &  $-1$ &    $-\frac{\mi }{r-Q_0}$
			\\  \hline
			$-\mi r-\mi Q_0$   &  $-1$      &  $-1$ &    $-\frac{\mi }{r+Q_0}$
			\\  \hline
		\end{tabular}
	\end{center}
\end{table}


Next, we define the associated eigenfunctions
\begin{equation}
	\label{hlj33}
	\begin{aligned}
		\bm{\psi}_{n}^{\pm}(t,z,\lambda) = (\mi r)^{-n} \mathbf{A}(\lambda) \bm{\phi}_{n}^{\pm}(t,z,\lambda) \mathbf{A}^{-1}(\lambda) \mathbf{J}^{-n}(z,\lambda) e^{-\bm{\Omega}(z,\lambda) t},
	\end{aligned}
\end{equation}
with $\mathbf{A}(\lambda) = \diag(1, \lambda)$. Immediately, Eq.~\eref{Laxpairjvzhenxingshiaa} implies that $\bm{\psi}_{n}^{\pm}(t,z,\lambda)$ satisfy the difference equations:
\begin{equation}
	\label{hlj38a}
	\begin{aligned}
		&\mi r \psi_{n+1,1}^{\pm}(t,z,\lambda)  =
		\begin{pmatrix}
			\frac{z}{\lambda} & \frac{Q_{n}(t)}{\lambda^2} \\
			\sigma  Q^*_{-n}(t) & \frac{1}{\lambda z} \\
		\end{pmatrix} \psi_{n,1}^{\pm}(t,z,\lambda),   \\
		&\mi r \psi_{n+1,2}^{\pm}(t,z,\lambda)  =
		\begin{pmatrix}
			z\lambda & Q_{n}(t) \\
			\sigma \lambda^2 Q^*_{-n}(t) & \frac{\lambda}{z} \\
		\end{pmatrix} \psi_{n,2}^{\pm}(t,z,\lambda).
	\end{aligned}
\end{equation}
Since the arguments $z, \lambda$ can be replaced by $\zeta$ through the relations in Eq.~\eref{hlj30}, the associated eigenfuncitons are thus denoted as $\bm{\psi}^{\pm}_n(t,\zeta)$ in the following.
According to Eq.~\eref{hlj21}, we have the following asymptotic behavior for $\bm{\psi}^{\pm}_n(t,\zeta)$:
\begin{equation}
	\label{hlj2.33}
	\begin{aligned}
		\bm{\psi}_{n}^{\pm}(t,\zeta)  \to \hat{\mathbf{Y}}_{\pm}(t,\zeta) :=  \begin{pmatrix}
			Q_{\pm} &  \mi r- \frac{1}{\zeta}   \\
			\zeta - \mi r   &  - \sigma Q^*_{\mp} \\
		\end{pmatrix}, \quad n \to \pm \infty.
	\end{aligned}
\end{equation}


On the other hand, substituting Eq.~\eref{hlj33} into Eq.~\eref{hlj26} and using Eq.~\eref{hlj30}, we have the relation between $\bm{\psi}_n^-$ and $\bm{\psi}_n^+$:

\begin{equation}
	\label{2.33}
	\begin{aligned}
		\bm{\psi}_{n}^{-}(t,\zeta) =  \bm{\psi}_{n}^{+}(t,\zeta)  \mathbf{T}(\zeta),
	\end{aligned}
\end{equation}
where $\mathbf{T}(\zeta)= (t_{ij})_{1\leq i,j \leq 2} = e^{\bm{\Omega}(\zeta)  t} \mathbf{J}^{n}(\lambda)  \mathbf{A}(\lambda)  \mathbf{S}(z,\lambda) \mathbf{A}^{-1}(\lambda) \mathbf{J}^{-n}(\lambda)  e^{-\bm{\Omega}(\zeta)  t}$.
Eq.~\eref{2.33} can be written in the column-wise form
\begin{equation}
	\label{hlj2.35}
	\begin{aligned}
		&\psi^{-}_{n,1}(t,\zeta) = t_{11}(\zeta) \psi^{+}_{n,1}(t,\zeta) +  t_{21}(\zeta)  \psi^{+}_{n,2}(t,\zeta),  \\
		&\psi^{-}_{n,2}(t,\zeta) =t_{12}(\zeta)    \psi^{+}_{n,1}(t,\zeta) +  t_{22}(\zeta) \psi^{+}_{n,2}(t,\zeta),
	\end{aligned}
\end{equation}
with
\begin{equation}
	\label{ss21}
	\begin{aligned}
		&t_{11}(\zeta) = s_{11}(z,\lambda), \quad
		t_{12}(\zeta) = \lambda^{2n-1} s_{12}(z,\lambda) e^{\Omega_{11}(\zeta) t- \Omega_{22}(\zeta) t}, \quad\\
		&t_{21}(\zeta) = \lambda^{-2n+1} s_{21}(z,\lambda) e^{\Omega_{22}(\zeta) t - \Omega_{11}(\zeta) t},\quad
		t_{22}(\zeta) = s_{22}(z,\lambda).
	\end{aligned}
\end{equation}
Eq.~\eref{ss21} implies that the following four functions can be expressed in terms of $\zeta$:
\begin{equation}
	\label{ss22}
	\begin{aligned}
		&s_{11}(z,\lambda) = s_{11}(\zeta), \quad
		\lambda^{-1} s_{12}(z,\lambda): = \overline{s}_{12}(\zeta), \quad\\
		&\lambda s_{21}(z,\lambda): = \overline{s}_{21}(\zeta),\quad
		s_{22}(z,\lambda) = s_{22}(\zeta).
	\end{aligned}
\end{equation}
By using Eqs.~\eref{hlj24} and~\eref{hlj33} together with Eq.~\eref{ss21}, we have
\begin{equation}
	\label{hlj4.20}
	\begin{aligned}
		&|\psi^{+}_{n,1}(t,\zeta), \psi^{+}_{n,2}(t,\zeta)| = - \frac{\mi r (\zeta + 1/\zeta - 2\mi r) }{\Delta_{n}(t)} , \\
		&|\psi^{-}_{n,1}(t,\zeta), \psi^{-}_{n,2}(t,\zeta)| = -  \frac{ \mi r (\zeta + 1/\zeta - 2\mi r) c_{\infty}}{\Delta_{n}(t)},
	\end{aligned}
\end{equation}
with $\Delta_{n}(t) = \displaystyle\prod_{k = n}^{+ \infty}  \frac{\sigma Q^*_{-k}(t) Q_{k}(t) - 1}{Q^2_{0} - 1 }$ and $c_{\infty} = \displaystyle\prod_{k = - \infty}^{ \infty} \frac{\sigma Q^*_{-k}(t) Q_{k}(t) - 1 }{Q^2_{0}  - 1 }$.
Also, the scattering coefficients can be written as the following $\zeta$-function:
\begin{equation}
	\label{hlj4.24}
	\begin{aligned}
		&s_{11}(\zeta) =  -\Delta_{n}(t) \frac{|\psi^{-}_{n,1}(t,\zeta), \psi^{+}_{n,2}(t,\zeta)|}{\mi r(\zeta + \frac{1}{\zeta } -2 \mi r)},  \quad
		s_{22}(\zeta) =  -\Delta_{n}(t) \frac{|\psi^{+}_{n,1}(t,\zeta), \psi^{-}_{n,2}(t,\zeta)|}{\mi r(\zeta + \frac{1}{\zeta } -2 \mi r)},\\
		&\overline{s}_{21}(z,\lambda)  =  -\Delta_{n}(t) \lambda^{-2n} \frac{|\psi^{+}_{n,1}(t,\zeta), \psi^{-}_{n,1}(t,\zeta)|}{\mi r(\zeta + \frac{1}{\zeta } -2 \mi r)},  \quad
		\overline{s}_{12}(z,\lambda)  =  -\Delta_{n}(t) \lambda^{2n}  \frac{|\psi^{-}_{n,2}(t,\zeta), \psi^{+}_{n,2}(t,\zeta)|}{\mi r(\zeta + \frac{1}{\zeta } -2 \mi r)}.
	\end{aligned}
\end{equation}

To further remove the asymptotic behavior of $\bm{\psi}^{\pm}_{n}(t,\zeta)$ as $n \to \infty$, we introduce the modified associated eigenfunciton
\begin{equation}
	\label{hlj5.17}
	\begin{aligned}
		\tilde{\bm{\psi}}^{\pm}_{n}(t,\zeta) & = \hat{\mathbf{Y}}^{-1}_{\pm}(t,\zeta)  \bm{\psi}^{\pm}_{n}(t,\zeta),
	\end{aligned}
\end{equation}
which satisfies $\displaystyle\lim_{n \to \pm \infty} \tilde{\bm{\psi}}_n = \mathbf{I}$.
Then, substituting Eq.~\eref{hlj5.17} into Eq.~\eref{hlj38a} yields
\begin{equation}
	\label{hlj57a}
	\begin{aligned}
		\tilde{\bm{\psi}}^{\pm}_{n+1,1}(t,\zeta)
		& = \left( \left(
		\begin{array}{ccc}
			1  &  0  \\
			0   &  \frac{1}{\lambda^{2}} \\
		\end{array}
		\right)  + \frac{\tilde{\mathbf{P}}^{\pm}_n(t,\zeta) }{\mi r \lambda}
		\right)
		\tilde{\bm{\psi}}^{\pm}_{n,1}(t,\zeta) , \\
		\tilde{\bm{\psi}}^{\pm}_{n+1,2}(t,\zeta)
		& = \left(
		\left(
		\begin{array}{ccc}
			\lambda^{2}  &  0  \\
			0   &  1 \\
		\end{array}
		\right) + \frac{\lambda \tilde{\mathbf{P}}^{\pm}_n(t,\zeta) }{\mi r}
		\right)
		 \tilde{\bm{\psi}}^{\pm}_{n,2}(t,\zeta),
	\end{aligned}
\end{equation}
with $\tilde{\mathbf{P}}^{\pm}_n(t,\zeta)  = \hat{\mathbf{Y}}^{-1}_{\pm}(z) \mathbf{A}(\lambda) \mathbf{U}_{n}(t,z) \mathbf{A}^{-1}(\lambda) \hat{\mathbf{Y}}_{\pm}(z)$, in which all the elements are the even functions of $\lambda$ and $z$.

Inversely, solving Eq.~\eref{hlj5.17} gives
\begin{equation}
	\label{xcx2.40}
	\begin{aligned}
			&\psi^{\pm}_{n,11}(t,\zeta) = Q_{\pm} \tilde{\psi}^{\pm}_{n,11}(t,\zeta) + \big(\mi r  - \frac{1}{\zeta} \big) \tilde{\psi}^{\pm}_{n,21}(t,\zeta),\quad
			\psi^{\pm}_{n,21}(t,\zeta) = (\zeta - \mi r)  \tilde{\psi}^{\pm}_{n,11}(t,\zeta) + Q^*_{\mp} \tilde{\psi}^{\pm}_{n,21}(t,\zeta),\\
			&\psi^{\pm}_{n,12}(t,\zeta) = Q_{\pm}\tilde{\psi}^{\pm}_{n,12}(t,\zeta) + \big(\mi r  - \frac{1}{\zeta} \big) \big) \tilde{\psi}^{\pm}_{n,22}(t,\zeta),\quad
			\psi^{\pm}_{n,22}(t,\zeta) = (\zeta - \mi r)  \tilde{\psi}^{\pm}_{n,12}(t,\zeta) + Q^*_{\mp} \tilde{\psi}^{\pm}_{n,22}(t,\zeta),
		\end{aligned}
\end{equation}
based on which we can repeat the proof of Theorem~\ref{th1}, and derive the analyticity of the columns of $\bm{\psi}^{\pm}_{n}(t,\zeta)$ as follows:
\begin{theorem}
	\label{th2}
	Under the condition $\Delta \mathbf{Q}_{n}^{\pm}(t) \in \Big\{ f_n \big| \displaystyle\sum_{j = \mp \infty}^{n}  | f_{j} | < \infty, \forall n \in \mathbb{Z} \Big\}$, the analyticity of the columns of $\bm{\psi}^{\pm}_{n}(t,\zeta)$ can be extended into the following regions:
	\begin{equation}
		\begin{aligned}
			&\psi^{-}_{n,1}(t,\zeta) : D^{-}, \quad \psi^{-}_{n,2}(t,\zeta): D^{+}, \\
			&\psi^{+}_{n,2}(t,\zeta) : D^{-}\backslash \{ \tfrac{1}{\mi r} \}, \quad \psi^{+}_{n,1}(t,\zeta) : D^{+} \backslash \{ \mi r \}.
		\end{aligned}
	\end{equation}
\end{theorem}

\begin{figure}[H]
	\centering
		\includegraphics[width=2.9in]{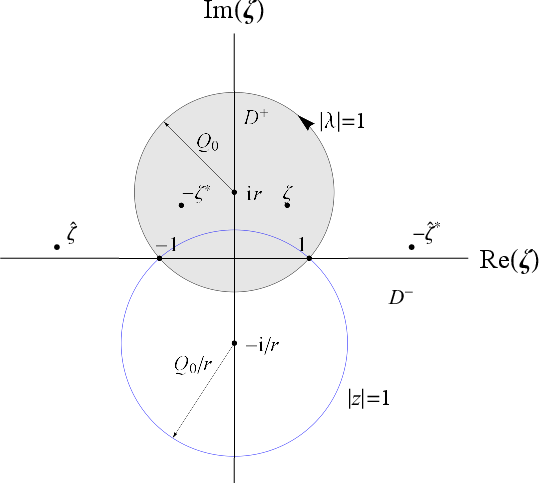}\hfill
	\caption{\small
		The regions $D^+$ (gray) and $D^-$ (white) in the complex $\zeta$-plane, which correspond to $|\lambda| < 1$ and $|\lambda| > 1$ in the complex $\lambda$-plane. The symmetries of  discrete eigenvalues are also displayed.  \label{f0} }
\end{figure}

\subsection{$\zeta$-asymptotic behavior of the associated eigenfunctions and scattering coefficients}
\label{Asymptotic behavior}
 From Eq.~\eref{hlj30}, it is important to point that
\begin{equation}
	\begin{aligned}
		(z, \lambda) \to 0 \Leftrightarrow \zeta \to \mi r,\quad (z, \lambda) \to \infty  \Leftrightarrow   \zeta \to \frac{1}{\mi r}.
	\end{aligned}
\end{equation}
Consequently, in the complex $\zeta$-plane, the points $\mi r$ and $\frac{1}{\mi r}$ play the role analogous to that of $0$ and $\infty$ in the complex
 $z$- and $\lambda$-planes.
Based on Eqs.~\eref{hlj29} amd~\eref{hlj30}, we present the asymptotic behaviors of $z^2$, $\lambda^2$ and $z\lambda$ at some distinguished points of $\zeta$ in Table~\ref{xcx4.61}.

\begin{table}[h]
	\small
	\caption{Asymptotic behaviors of $z^2$, $\lambda^2$ and $z\lambda$ at $\zeta \to 0, \infty, \mi r$ and $\frac{1}{\mi r}$.  \label{xcx4.61}}
	\begin{center}
		\begin{tabular}{c|c|c|c|c}
			\hline
			  &   $\zeta  \to 0$ & $\zeta  \to \infty$ & $\zeta  \to \mi r$  & $\zeta  \to \frac{1}{\mi r}$
			\\  \hline
			$z^2$   &  $\frac{\mi r}{\zeta}$ &   $\frac{1}{\mi r \zeta}$  &  $\frac{\mi (\zeta-\mi r )}{r Q^2_0 } $  & $\frac{Q^2_0 }{\mi r (\zeta - \frac{1}{\mi r})} $
			\\  \hline
			$\lambda^2$   &  $\mi r\zeta$ &   $\frac{\zeta}{\mi r} $  &  $-\frac{\mi r (\zeta-\mi r )}{Q^2_0  }$      & $\frac{Q^2_0  }{(\mi r)^{3} (\zeta - \frac{1}{\mi r})}$
			\\  \hline
			$z\lambda$    &  $\mi r$ &   $\frac{1}{\mi r}$  &  $-\frac{\zeta-\mi r }{  Q^2_0  }$     & $\frac{Q^2_0 }{\mi r^{2} (\zeta - \frac{1}{\mi r}) } $
			\\  \hline
		\end{tabular}
	\end{center}
\end{table}


Next, we study the asymptotic behavior of associated eigenfunctions $\psi_n^{\pm}$ as $\zeta \to 0, \infty, \mi r$ and $\frac{1}{\mi r}$. As an illustrative example, we demonstrate the analysis for the case $\zeta \to \mi r$.
Based on the analyticity of $\psi_{n,1}^{+}$ and $\psi_{n,2}^{-}$ at $\zeta = \mi r$, we formally expand the two components of $\psi_{n,1}^{+}$ and $\psi_{n,2}^{-}$ as follows:
\begin{subequations}
	\label{hlj79}
	\begin{align}
			& \psi_{n,11}^{+}(t,\zeta) = \psi_{n,11}^{+,0}(t)  + \psi_{n,11}^{+,1}(t) \big(\zeta - \mi r \big) + \cdots,\\
			& \psi_{n,21}^{+}(t,\zeta) = \psi_{n,21}^{+,1}(t) \big(\zeta - \mi r \big) + \psi_{n,21}^{+,1}(t) \big(\zeta - \mi r \big)^2 + \cdots, \\
			& \psi_{n,12}^{-}(t,\zeta) = \psi_{n,12}^{-,0}(t) + \psi_{n,12}^{-,1}(t) \big(\zeta - \mi r \big) + \cdots, \\
			& \psi_{n,22}^{-}(t,\zeta) = \psi_{n,22}^{-,0}(t) + \psi_{n,22}^{-,1}(t) \big(\zeta - \mi r \big) + \cdots,
	\end{align}
\end{subequations}
where $\psi_{n,ij}^{\pm}(t,\zeta)$ is the $(i,j)$-element of $\psi_{n}^{\pm}(t,\zeta)$, and $\psi_{n,ij}^{\pm,k}(t)$ are the expansion coefficients. Inserting Eqs.~\eref{hlj79} into Eq.~\eref{hlj38a} and comparing the coefficients of the constant terms, we obtain the following relations:
\begin{subequations}
	\label{hlj92}
	\begin{align}
		&\mi r \psi_{n+1,11}^{+,0}(t) = \frac{1}{\mi r} \psi_{n,11}^{+,0}(t)   + \frac{\mi Q_{n}(t)  Q^2_0 }{r} \psi_{n,21}^{+,1}(t),  \label{hlj92a} \\
		&\sigma Q^*_{-n}(t) \psi_{n,11}^{+,0}(t) -  Q^2_0  \psi_{n,21}^{+,1}(t) = 0,  \label{hlj92b} \\
		&\mi r \psi_{n+1,12}^{-,0}(t) = Q_{n}(t) \psi_{n,22}^{-,0}(t),   \label{hlj92c} \\
		&\psi_{n+1,22}^{-,0}(t) = \psi_{n,22}^{-,0}(t).   \label{hlj92d}
	\end{align}
\end{subequations}
Looking at the asymptotic behavior of $\psi_n^{\pm}$ in Eq.~\eref{hlj2.33}, we derive from Eqs.~\eref{hlj92a}-\eref{hlj92d} that
\begin{equation}
	\label{hlj2.48}
	\begin{aligned}
		&\psi_{n,12}^{-,0}(t) = - \frac{\sigma Q^{*}_{+} Q_{n-1}(t)}{\mi r}, \quad
		\psi_{n,22}^{-,0}(t)  = - \sigma Q^{*}_{+},    \\
		& \psi_{n,11}^{+,0}(t) = \frac{Q_{+}}{\Delta_{n}(t)}, \quad
		\psi_{n,21}^{+,1}(t) = \frac{Q_{+}}{\Delta_{n}(t)} \frac{ \sigma Q^*_{-n}(t)  }{Q^2_0}
	\end{aligned}
\end{equation}
Therefore, the asymptotic behavior of associated eigenfunctions as $\zeta \to \mi r$ can be obtained as:
\begin{equation}
	\label{hlj2.491}
	\begin{aligned}
		&\psi_{n,1}^{+}(t) \to  \frac{Q_+}{\Delta_n(t)} \left(
		\begin{array}{ccc}
			1 \\
			\frac{\sigma Q^*_{-n}(t)}{Q^2_0}  \big(\zeta - \mi r \big)   \\
		\end{array}
		\right), \quad
		\psi_{n,2}^{-}(t) \to -\sigma Q^*_+  \left(
		\begin{array}{ccc}
			\frac{Q_{n-1}(t)}{\mi r}  \\
			1  \\
		\end{array}
		\right), \quad   \zeta \to \mi r.
	\end{aligned}
\end{equation}

By repeating the above process, we obtain
\begin{subequations}
	\label{hlj2.49}
	\begin{align}
	&\psi_{n,1}^{-}(t) \to  \left(
	\begin{array}{ccc}
		Q_{n-1}(t) \\
		\zeta   \\
	\end{array}
	\right), \quad
	\psi_{n,2}^{+}(t) \to  \frac{1}{\Delta_n(t)} \left(
	\begin{array}{ccc}
		\mi r  \\
		-\sigma Q^*_{-n}(t)  \\
	\end{array}
	\right),
	\quad \quad \quad  \quad \quad \quad \quad  \quad   \zeta \to \infty,\\
	&\psi_{n,1}^{+}(t) \to  \frac{1}{\Delta_n(t)} \left(
	\begin{array}{ccc}
		Q_{n}(t) \\
		-\mi r   \\
	\end{array}
	\right), \quad
	\psi_{n,2}^{-}(t) \to  \left(
	\begin{array}{ccc}
		-\frac{1}{\zeta}  \\
		-\sigma Q^*_{-n+1}(t)  \\
	\end{array}
	\right),
	\quad \quad \quad  \quad \quad \quad \quad  \quad   \zeta \to 0,\\
	&\psi_{n,1}^{-}(t) \to  Q_{-}\left(
	\begin{array}{ccc}
		1   \\
		\frac{\sigma Q^*_{-n+1}(t)}{\mi r}   \\
	\end{array}
	\right), \quad
	\psi_{n,2}^{+}(t) \to  - \frac{\sigma Q^*_{-}}{\Delta_n(t)} \left(
	\begin{array}{ccc}
		\frac{Q_n(t) r^2}{Q^2_0}  \big(\zeta - \frac{1}{\mi r}\big)  \\
		1  \\
	\end{array}
	\right), \quad \quad \quad    \zeta \to \frac{1}{\mi r}.
	\end{align}
\end{subequations}
In addition, based on Eq.~\eref{hlj4.24}, the asymptotic behavior of scattering coefficients can be derived as follows:
\begin{equation}
	\label{hlj2.51}
	\begin{aligned}
		&s_{11}(\zeta) =
		\left\{
		\begin{aligned}
			&1,                                 \quad \quad \quad \quad  \quad   \, \, \,    \zeta \to \infty, \\
			&e^{\mi (\theta_{-}- \theta_{+})},  \quad\quad \zeta \to   \frac{1}{\mi r}, \\
		\end{aligned}
		\right. \\
		&s_{22}(\zeta) =
		\left\{
		\begin{aligned}
			&1,                                 \quad \quad \quad \quad  \quad    \,\,  \,  \zeta \to 0, \\
			&e^{\mi (\theta_{-}- \theta_{+})},  \quad\quad \zeta \to   \mi r. \\
		\end{aligned}
		\right.
	\end{aligned}
\end{equation}

\subsection{Symmetries of eigenfunctions and scattering matrix}
\label{2.5}

In this part, we define $\hat{\zeta}: = \frac{1}{z\lambda} =  \frac{\mi r \zeta -1}{\zeta - \mi r} $ and omit the $t$-dependence for the involved functions.
Owing to Eqs.~\eref{hlj17} and~\eref{hlj29},
both $\bm{\phi}_n(z,\lambda,\zeta)$ and $\bm{\phi}_n(z,\frac{1}{\lambda},\hat{\zeta})$ are regarded as the solutions to the Lax pair~\eref{hlj9a}.
Considering the asymptotic behavior of $\bm{\phi}_n(z,\lambda,\zeta)$ in Eq.~\eref{hlj21} as $n \to \pm \infty$ and using the relation $(\mi r \lambda - z)(z - \frac{\mi r}{\lambda}) = Q_0^2$, we can derive the following symmetry.
\begin{lemma}
	\label{le2.17}
	If $\bm{\phi}_n(z,\lambda,\zeta)$ solves the Lax pair~\eref{hlj9a},  $\bm{\phi}_n(z,\frac{1}{\lambda},\hat{\zeta})$ is the solution of Eq.~\eref{hlj9a}. Moreover, the eigenfunctions and scattering matrix possess the  symmetry as follows:
	\begin{subequations}
		\begin{align}
			&\bm{\phi}^{\pm}_{n}(z,\frac{1}{\lambda},\hat{\zeta}) = \bm{\phi}^{\pm}_{n}(z,\lambda,\zeta) \bm{\Pi}_{\pm}(z,\lambda),
			\quad
			\bm{\Pi}_{\pm}(z,\lambda) = \begin{pmatrix}
				0  &  \frac{ \frac{\mi r}{\lambda} - z}{ Q_{\pm}}      \\
				\frac{ \frac{\mi r}{\lambda} - z}{-\sigma Q^*_{\mp}}  &  0 \\
			\end{pmatrix},
			\label{xcx56} \\
			&\mathbf{S}(z,\lambda,\zeta) = \mathbf{\Pi}_{+}(z) \mathbf{S}(z,\frac{1}{\lambda},\hat{\zeta}) \mathbf{\Pi}^{-1}_{-}(z). \label{hlj2.53}
		\end{align}
	\end{subequations}
\end{lemma}

On the other hand, referring to the work in Ref.~\cite{Ablowitz2020}, we derive the `backward' scattering problem by inverting Eq.~\eref{Laxpairjvzhenxingshiaa}:
\begin{equation}
	\label{hlj2.2}
	\begin{aligned}
		& \bm{w}_{n-1}(z)
		= \left(
		\begin{array}{ccc}
			\frac{1}{z} & - Q_{n}(t) \\
			- \sigma Q^*_{-n}(t) & z \\
		\end{array}
		\right) \bm{w}_{n}(z),
	\end{aligned}
\end{equation}
where
\begin{equation}
	\label{hlj2.3}
	\begin{aligned}
		&\bm{w}_{n}(z) = \Upsilon_{n}(t) \bm{\phi}_{n+1}(t,z), \quad \Upsilon_{n}(t) =  \frac{1}{ (\mi r)^{2n} } \displaystyle\prod_{k = - \infty}^{n} \frac{1 - \sigma Q_+ Q^*_-}{1 - \sigma Q_k(t)  Q^*_{-k}(t) }.
	\end{aligned}
\end{equation}
Then, using Eqs.~\eref{hlj2.2},~\eref{hlj2.3} and the relation
\begin{equation}
	\begin{aligned}
		\bm{\sigma}_{1} \mathbf{U}^*_{-n}(z^*) \bm{\sigma}^{-1}_{1} = \left(
		\begin{array}{ccc}
			\frac{1}{z} &  - Q_{n}(t) \\
			-\sigma Q^*_{-n}(t)    & z \\
		\end{array}
		\right), \quad
		\bm{\sigma}_{1} = \left(
		\begin{array}{ccc}
			0 &  1 \\
			-\sigma &  0 \\
		\end{array}
		\right),
	\end{aligned}
\end{equation}
we can derive the second symmetry about $\bm{\phi}_n(z,\lambda,\zeta)$ and $\mathbf{S}(z,\lambda,\zeta)$.

\begin{lemma}
	\label{le2.17a}
	If $\bm{\phi}_n(z,\lambda,\zeta)$ solves the Lax pair~\eref{hlj9a}, $\bm{\sigma}_{1} \bm{\phi}^*_{-n} (z^*, -\lambda^*, -\zeta^*)  \bm{\sigma}^{-1}_{1}$ is the solution of Eq.~\eref{hlj2.2}. Moreover, the eigenfunctions and scattering matrix possess the symmetry as follows:
	\begin{subequations}
		\label{hlj106}
		\begin{align}
			&\bm{\sigma}_{1} \big( \bm{\phi}^{-}_{-n}(z^*, -\lambda^*, -\zeta^*) \big)^* \bm{\sigma}^{-1}_{1} = \Upsilon_{n}(t) \bm{\phi}^{+}_{n+1}(z,\lambda,\zeta) \mathbf{C}(\lambda), \quad
			\mathbf{C}(\lambda) = \begin{pmatrix}
				- \frac{\sigma c_{\infty}}{\mi r \lambda }   & 0   \\
				0  & -  \frac{ \sigma \lambda c_{\infty}}{\mi r }    \\
			\end{pmatrix}, \label{hlj106a}\\
		   &\mathbf{S}(z,\lambda,\zeta) = \bm{\sigma}_{1} \mathbf{C}(\lambda)  \mathbf{S}(z^*,-\lambda^*, -\zeta^*) \big(\bm{\sigma}_{1} \mathbf{C}(\lambda)\big)^{-1}. \label{hlj2.56}
		\end{align}
	\end{subequations}
\end{lemma}

\section{Inverse scattering problem}
\label{sec3}
In this section, we formulate and solve the RHP for the associated eigenfunctions $\bm{\psi}^{\pm}_{n}(t,\zeta)$, calculate the pole contributions, and obtain the potential reconstruction formula when the scattering coefficients have an arbitrary number of simple zeros.

\subsection{Riemann-Hilbert problem and pole contributions}
Recall that $\Big\{\frac{\psi^{-}_{n,1}(t,\zeta)}{s_{11}(\zeta)}, \psi^{+}_{n,2}(t,\zeta)\Big\}$ and $\Big\{\frac{\psi^{-}_{n,2}(t,\zeta)}{s_{22}(\zeta)}, \psi^{+}_{n,1}(t,\zeta)\Big\}$ are meromorphic in the regions $D^{-}$ and $D^{+}$, respectively. By virtue of Eq.~\eref{hlj2.35}, we can obtain the formulation of  the RHP as follows:
\begin{lemma}
	\label{le4.1}
	The meromorphic functions $\Big\{\frac{\psi^{-}_{n,1}(t,\zeta)}{s_{11}(\zeta)}, \psi^{+}_{n,1}(t,\zeta)\Big\}$ and $\Big\{\frac{\psi^{-}_{n,2}(t,\zeta)}{s_{22}(\zeta)}, \psi^{+}_{n,2}(t,\zeta)\Big\}$ satisfy jump conditions:
\begin{subequations}
	\label{hlj126}
	\begin{align}
		&\frac{\psi^{-}_{n,1}(t,\zeta)}{s_{11}(\zeta)}  - \psi^{+}_{n,1}(t,\zeta) = \rho(\zeta)       \lambda^{-2n}        \psi^{+}_{n,2}(t,\zeta) e^{\Omega_{22}(\zeta) t - \Omega_{11}(\zeta) t},  	\label{hlj126a}\\
		&\frac{\psi^{-}_{n,2}(t,\zeta)}{s_{22}(\zeta)}  - \psi^{+}_{n,2}(t,\zeta) = \bar{\rho}(\zeta) \lambda^{2n}         \psi^{+}_{n,1}(t,\zeta) e^{\Omega_{11}(\zeta) t - \Omega_{22}(\zeta) t}.	\label{hlj126b}
	\end{align}
\end{subequations}
with $\rho(\zeta)= \frac{\overline{s}_{21}(\zeta)}{s_{11}(\zeta)}$ and $\bar{\rho}(\zeta) = \frac{\overline{s}_{12}(\zeta)}{s_{22}(\zeta)}$ are the reflection coefficients.
Meanwhile, they have the following asymptotic behaviors:
\begin{subequations}
	\label{hlj3.2}
	\begin{align}
		&\frac{\psi^{-}_{n,1}(t,\zeta)}{s_{11}(\zeta)} \to
		\left(
		\begin{array}{ccc}
			Q_{n-1}(t) + o(1)\\
			\zeta   + O(1) \\
		\end{array}
		\right),
		&\zeta \to \infty, \\
		& \psi^{+}_{n,2}(t,\zeta) \to  \frac{1}{\Delta_n(t)} \left(
		\begin{array}{ccc}
			\mi r  \\
			-\sigma Q^*_{-n}(t)  \\
		\end{array}
		\right) + O\Big(\frac{1}{\zeta} \Big),
		& \zeta \to \infty, \label{hlj3.2a}\\
		&\frac{\psi^{-}_{n,1}(t,\zeta)}{s_{11}(\zeta)} \to
		Q_{-} e^{\mi (\theta_+ - \theta_{-})} \left(
		\begin{array}{ccc}
			1   \\
			\frac{\sigma Q^*_{-n+1}(t)  }{r}   \\
		\end{array}
		\right) + O\Big(\zeta - \frac{1}{\mi r}\Big),
		& \zeta \to \frac{1}{\mi r}, \\
		& \psi^{+}_{n,2}(t,\zeta) \to  -\sigma \frac{Q^*_{-}}{\Delta_n(t)} \left(
		\begin{array}{ccc}
			0  \\
			1  \\
		\end{array}
		\right) + O\Big(\zeta - \frac{1}{\mi r}\Big),
		& \zeta \to \frac{1}{\mi r},  \label{hlj3.2b} \\
		&\frac{\psi^{-}_{n,2}(t,\zeta)}{s_{22}(\zeta)} \to
		\left(
		\begin{array}{ccc}
			-\frac{1}{\zeta}  + O(1) \\
			-\sigma Q^*_{-n+1}(t)  + o(1) \\
		\end{array}
		\right), \quad
		& \zeta \to 0, \\
		&\psi_{n,1}^{+}(t,\zeta) \to  \frac{1}{\Delta_n(t)} \left(
		\begin{array}{ccc}
			Q_{n}(t) \\
			-\mi r   \\
		\end{array}
		\right) + O(\zeta),
		&\zeta \to 0, \label{hlj3.2c} \\
		&\frac{\psi^{-}_{n,2}(t,\zeta)}{s_{22}(\zeta)} \to
		- \sigma  Q^*_+ e^{\mi (\theta_+ - \theta_{-})}  \left(
		\begin{array}{ccc}
			\frac{Q_{n-1}(t)}{\mi r}  \\
			1  \\
		\end{array}
		\right)+ O(\zeta - \mi r),
		& \zeta \to \mi r,\\
		&\psi_{n,1}^{+}(t,\zeta) \to  \frac{Q_+}{\Delta_n(t)} \left(
		\begin{array}{ccc}
			1 \\
			0   \\
		\end{array}
		\right) + O(\zeta - \mi r),
		&  \zeta \to \mi r. \label{hlj3.2d}
	\end{align}
\end{subequations}
\end{lemma}
In fact, Eqs.~\eref{hlj126a} and~\eref{hlj126b} can be written in the following matrix form:
\begin{subequations}
	\label{hlj126aa}
	\begin{align}
		&\Bigg(\frac{\psi^{-}_{n,1}(t,\zeta)}{s_{11}(\zeta)}, \psi^{+}_{n,2}(t,\zeta)\Bigg) =
		\Bigg(\psi^{+}_{n,1}(t,\zeta), \frac{\psi^{-}_{n,2}(t,\zeta)}{s_{22}(\zeta)} \Bigg) \mathbf{H}_n(t,\zeta),
	\end{align}
\end{subequations}
where $\mathbf{H}_n(t,\zeta) = \left(
\begin{array}{ccc}
	1 -  \rho(\zeta) \bar{\rho}(\zeta)  & - \bar{\rho}(\zeta) \lambda^{2n} e^{\Omega_{11}(\zeta) t - \Omega_{22}(\zeta) t} \\
	\rho(\zeta) \lambda^{-2n}  e^{\Omega_{22}(\zeta) t - \Omega_{11}(\zeta) t} & 1 \\
\end{array}
\right)$ is called the jump matrix.

Next, we regularize the RHP~\eref{hlj126} by removing the asymptotic behaviors of $\mathbf{M}^{\pm}$ and the contributions at the zeros of scattering coefficients $s_{11}(\zeta)$, $s_{22}(\zeta)$.
Considering the analyticity of $s_{11}(\zeta)$ and $s_{22}(\zeta)$ in $D^{-}$ and $D^{+}$, respectively, we suppose that $s_{22}(\zeta)$ has simple zeros $\{\zeta_i\}^{n_1}_{i=1}$ in $D^{+}$. Then, based on the symmetries~\eref{hlj2.53} and~\eref{hlj2.56}, we know that $s_{22}(\zeta)$ also has simple zeros $\{-\zeta^*_i\}^{n_1}_{i=1}$, and $s_{11}(z)$ has simple zeros $\{\hat{\zeta}_i, -\hat{\zeta}^*_i \}^{n_1}_{i=1}$. Correspondingly, the pole contributions can be obtained from the residue conditions of $\frac{\psi^{-}_{n,2}(t,\zeta)}{s_{22}(\zeta)}$ and $\frac{\psi^{-}_{n,1}(t,\zeta)}{s_{11}(\zeta)}$ as follows:

\begin{lemma}
	\label{le4.2}
	At the discrete eigenvalues $\zeta = \zeta_i$ ($1 \leq i \leq n_1$),
	$\frac{\psi^{-}_{n,2}(t,\zeta)}{s_{22}(\zeta)}$ and $\frac{\psi^{-}_{n,1}(t,\zeta)}{s_{11}(\zeta)}$ possess the residue conditions:
\begin{equation}
	\label{hlj132}
	\begin{aligned}
		&\underset{\zeta = \zeta_i}{\Res} \Bigg[\frac{\psi^{-}_{n,2}(t,\zeta)}{s_{22}(\zeta)}   \Bigg]
		= C_n(t,\zeta_i) \psi^{+}_{n,1}(t,\zeta_i), \quad
		\underset{\zeta = -\zeta^*_i}{\Res} \Bigg[ \frac{\psi^{-}_{n,2}(t,\zeta)}{s_{22}(\zeta)}  \Bigg]
		= C_n(t,-\zeta^*_i) \psi^{+}_{n,1}(t,-\zeta^*_i), \,\, \\
		&\underset{\zeta = \hat{\zeta}_i}{\Res}  \Bigg[ \frac{\psi^{-}_{n,1}(t,\zeta)}{s_{11}(\zeta)}  \Bigg]
		= \bar{C}_n(t,\hat{\zeta}_i) \psi^{+}_{n,2}(t,\hat{\zeta}_i) , \quad
		\underset{\zeta = -\hat{\zeta}^*_i}{\Res} \Bigg[ \frac{\psi^{-}_{n,1}(t,\zeta)}{s_{11}(\zeta)}  \Bigg]
		= \bar{C}_n(t, -\hat{\zeta}^*_i) \psi^{+}_{n,2}(t,-\hat{\zeta}^*_i),
	\end{aligned}
\end{equation}
with
\begin{equation}
	\label{hlj133}
	\begin{aligned}
		&C_n(t,\zeta_i)  = \frac{\lambda^{2n}(\zeta_i) \overline{s}_{12}(\zeta_{i}) }{s'_{22}(\zeta_i)}  e^{\Omega_{11}(\zeta_{i}) t - \Omega_{22}(\zeta_{i}) t}, \quad
		C_n(t,-\zeta^*_i) = \frac{\lambda^{2n}(-\zeta^*_i) \overline{s}_{12}(-\zeta^*_{i}) }{s'_{22}(-\zeta^*_i)}  e^{\Omega_{11}(-\zeta^*_i) t- \Omega_{22}(-\zeta^*_i) t}, \,\, \\
		&\bar{C}_n(t,\hat{\zeta}_i) = \frac{\lambda^{-2n}(\hat{\zeta}_i)  \overline{s}_{21}(\hat{\zeta}_i) }{s'_{11}(\hat{\zeta}_i)} e^{\Omega_{22}(\hat{\zeta}_i) t- \Omega_{11}(\hat{\zeta}_i) t} , \quad
		\bar{C}_n(t, -\hat{\zeta}^*_i) = \frac{\lambda^{-2n}(-\hat{\zeta}^*_i)  \overline{s}_{21}(-\hat{\zeta}^*_i) }{s'_{11}(-\hat{\zeta}^*_i)}  e^{\Omega_{22}(-\hat{\zeta}^*_i) t- \Omega_{11}(-\hat{\zeta}^*_i) t},
	\end{aligned}
\end{equation}
and
\begin{equation}
	\label{hlj3.14}
	\begin{aligned}
		\overline{s}_{12}(\zeta_i)  = - \sigma \frac{Q_+^*}{Q_+ } \overline{s}_{21}(\hat{\zeta}_i),\quad
		\overline{s}_{12}(\zeta_{i}) \overline{s}^*_{12}(-\zeta^*_{i})
		= \sigma \Big(\frac{\mi r \zeta_{i} -1 }{\zeta_{i}(\zeta_{i} - \mi r)}\Big)^2
		c_{\infty} , \quad
		\overline{s}_{21}(\hat{\zeta}_{i}) \overline{s}^*_{21}(-\hat{\zeta}^*_{i})
		= \sigma
		\Big(\frac{ \hat{\zeta}_i (\hat{\zeta}_i - \mi r)}{\mi r \hat{\zeta}_i -1}\Big)^2
		c_{\infty}.
	\end{aligned}
\end{equation}
\end{lemma}
The proof for the constraints on norming constants in Eq.~\eref{hlj3.14} can be seen in Appendix~\ref{appendixA3a}.

\subsection{Trace formula}
Based on the analyticity of $s_{11}(z)$ and $s_{22}(z)$, we define
\begin{equation}
	\label{sl39}
	\begin{aligned}
		&A(\zeta) = s_{11}(\zeta) \displaystyle\prod_{i = 1}^{n_1} \frac{ \zeta - \zeta_i  }{\zeta- \hat{\zeta}_i }   \frac{ \zeta + \zeta^*_i }{\zeta + \hat{\zeta}^*_i },  \quad
		B(\zeta) = s_{22}(\zeta) \displaystyle\prod_{i = 1}^{n_1} \frac{ \zeta - \hat{\zeta}_i  }{\zeta- \zeta_i }   \frac{ \zeta + \hat{\zeta}^*_i }{\zeta + \zeta^*_i }.
	\end{aligned}
\end{equation}
Resultingly, $A(\zeta)$ and $B(\zeta)$ are, respectively, analytic in $D^{-}$ and $D^{+}$, and they have no zeros in the corresponding regions. According to Eq.~\eref{hlj2.25}, we can obtain the following RHP about $s_{11}(z)$ and $s_{22}(z)$:
\begin{equation}
	\begin{aligned}
		&\ln \big(s_{11}(\zeta)\big) + \ln \big( s_{22}(\zeta)\big)    = \ln \big( c_{\infty} + s_{12}(\zeta) s_{21}(\zeta) \big).
	\end{aligned}
\end{equation}
By virtue of Plemelj's formula, we can obtain the trace formula
\begin{equation}
	\label{hlj3.7}
	\begin{aligned}
		&s_{11}(\zeta)  = \displaystyle\prod_{i = 1}^{n_1} \frac{ \zeta - \hat{\zeta}_i  }{\zeta- \zeta_i }   \frac{ \zeta + \hat{\zeta}^*_i }{\zeta + \zeta^*_i }
		\exp\Big[  - \frac{1}{2 \pi \mathrm{i}}  \int_{ \Sigma }^{ }  \frac{\ln \big( c_{\infty} + s_{12}(\zeta) s_{21}(\zeta) \big) }{v- \zeta}dv \Big],
		\quad \zeta \in D^{-}, \\
		&s_{22}(\zeta)  = \displaystyle\prod_{i = 1}^{n_1} \frac{ \zeta - \zeta_i  }{\zeta- \hat{\zeta}_i }   \frac{ \zeta + \zeta^*_i }{\zeta + \hat{\zeta}^*_i }
		\exp\Big[\frac{1}{2 \pi \mathrm{i}}  \int_{ \Sigma }^{ }  \frac{\ln \big( c_{\infty} + s_{12}(\zeta) s_{21}(\zeta)    \big) }{v - \zeta}dv \Big].
		\quad \zeta \in D^{+}.
	\end{aligned}
\end{equation}

\subsection{Reconstruction formula}
In view that $\psi^{+}_{n,1}(t,\zeta)$ has the pole $\zeta = \mi r$ while $\psi^{+}_{n,2}(t,\zeta)$ has the pole $\zeta =\frac{1}{ \mi r}$, we divide Eqs.~\eref{hlj126a} and~\eref{hlj126b} respectively by $\zeta - \mi r$ and $\zeta - \frac{1}{\mi r}$, and then obtain the modified RHP as follows:
\begin{subequations}
	\label{hlj135}
	\begin{align}
		&\frac{\psi^{-}_{n,1}(t,\zeta)}{(\zeta - \mi r)s_{11}(\zeta)}  = \frac{\psi^{+}_{n,1}(t,\zeta)}{\zeta - \mi r} +
		\rho(\zeta) \frac{\lambda^{-2n}}{\zeta - \mi r} \psi^{+}_{n,2}(t,\zeta),\label{hlj135a1} \\
		&\frac{\psi^{-}_{n,2}(t,\zeta)}{(\zeta - \frac{1}{\mi r})s_{22}(\zeta)}  = \frac{\psi^{+}_{n,2}(t,\zeta)}{\zeta - \frac{1}{\mi r}}
		+ \bar{\rho}(\zeta) \frac{ \lambda^{2n}}{\zeta - \frac{1}{\mi r}}  \psi^{+}_{n,1}(t,\zeta),\label{hlj135b1}
	\end{align}
\end{subequations}
where the left-hand sides are analytic in $D^{-}$ and $D^{+}$, respectively.
By subtracting the asymptotic behavior and pole contributions from both sides of Eqs.~\eref{hlj135a1} and~\eref{hlj135b1}, we solve the modified RHP~\eref{hlj135}, yielding:
\begin{equation}
	\label{hlj139}
	\begin{aligned}
		\psi^{+}_{n,1}(t,\zeta) =
		&
		\left(
		\begin{array}{ccc}
			\frac{Q_+}{\Delta_n(t)}  \\
			\zeta - \mi r  \\
		\end{array}
		\right)
		+
		\displaystyle\sum_{i = 1}^{n_1}
		\Bigg(
		\frac{(\zeta - \mi r)
			\bar{C}_n(t,\hat{\zeta}_i) \psi^{+}_{n,2}(t,\hat{\zeta}_i)
		}{(\hat{\zeta}_i - \mi r)(\zeta - \hat{\zeta}_i)} +
		\frac{(\zeta - \mi r)
			\bar{C}_n(t,-\hat{\zeta}^*_i) \psi^{+}_{n,2}(t,-\hat{\zeta}^*_i)
		}{(-\hat{\zeta}^*_i -\mi r)(\zeta+\hat{\zeta}^*_i)}
		\Bigg) \\
		&
		-  \frac{1}{2 \pi \mi}  \int_{w\in \Sigma} \frac{(\zeta - \mi r)\lambda^{-2n}(w) }{(w -\mi r)(w -\zeta)} \rho(w)  \psi^{+}_{n,2}(t,w) dw,
	\end{aligned}
\end{equation}
and
\begin{equation}
	\label{hlj139a}
	\begin{aligned}
		\psi^{+}_{n,2}(t,\zeta) =
		&\left(
		\begin{array}{ccc}
			\mi r - \frac{1}{\zeta}  \\
			- \frac{\sigma Q^*_{-}}{\Delta_n(t)} \\
		\end{array}
		\right)
		+
		\displaystyle\sum_{i = 1}^{n_1}
		\Bigg(
		\frac{(\zeta - \frac{1}{\mi r})
			C_n(t,\zeta_i) \psi^{+}_{n,1}(t,\zeta_i)
		}{(\zeta_i - \frac{1}{\mi r})(\zeta - \zeta^*_i)} +
		\frac{(\zeta - \frac{1}{\mi r})
			 C_n(t,-\zeta^*_i) \psi^{+}_{n,1}(t,-\zeta^*_i)
		}{(-\zeta^*_i - \frac{1}{\mi r})(\zeta + \zeta^*_i)}
		\Bigg) \\
		&
		- \frac{1}{2 \pi \mi}   \int_{w\in \Sigma} \frac{(\zeta - \frac{1}{\mi r}) \lambda^{2n}(w) }{(w -  \frac{1}{\mi r})(w-\zeta)}   \bar{\rho}(w)     \psi^{+}_{n,1}(t,w)dw. \\
	\end{aligned}
\end{equation}
Meanwhile, imposing the asymptotic behavior of $\psi^{+}_{n,1}(t,\zeta)$ as $\zeta \to 0$ and that of $\psi^{+}_{n,2}(t,\zeta) $ as $\zeta \to \infty$
(see Eqs.~\eref{hlj3.2a} and~\eref{hlj3.2c}) on Eqs.~\eref{hlj139} and~\eref{hlj139a} gives
\begin{subequations}
	\label{hlj3.22}
	\begin{align}
		\frac{1}{\Delta_n(t)} \left(
		\begin{array}{ccc}
			Q_{n}(t) \\
			-\mi r   \\
		\end{array}
		\right)
		= &
		\left(
		\begin{array}{ccc}
			\frac{Q_+}{\Delta_n(t)}  \\
			- \mi r  \\
		\end{array}
		\right)
		+
		\displaystyle\sum_{i = 1}^{n_1}
		\Big(
		\frac{\mi r \bar{C}_n (t, \hat{\zeta}_i) \psi^{+}_{n,2}(t,\hat{\zeta}_{i})     }{(\hat{\zeta}_i -\mi r)\hat{\zeta}_i} -
		\frac{\mi r \bar{C}_n (t, -\hat{\zeta}_i^*) \psi^{+}_{n,2}(t,-\hat{\zeta}^*_{i}) }{(-\hat{\zeta}^*_i -\mi r)\hat{\zeta}^*_i }
		\Big)\nonumber \\
		&
		+  \frac{1}{2 \pi \mi}  \int_{w\in \Sigma} \frac{\mi r \lambda^{-2n}(w) }{w (w -\mi r)} \rho(w)  \psi^{+}_{n,2}(t,w)dw,  \label{hlj3.22a} \\
		\frac{1}{\Delta_n(t)} \left(
		\begin{array}{ccc}
			\mi r  \\
			-\sigma Q^*_{-n}(t)  \\
		\end{array}
		\right)
		=&\left(
		\begin{array}{ccc}
			\mi r  \\
			-\frac{\sigma Q^*_{-}}{\Delta_n(t)} \\
		\end{array}
		\right)
		+
		\displaystyle\sum_{i = 1}^{n_1}
		\Big(
		\frac{ C_n (t, \zeta_i) \psi^{+}_{n,1}(t,\zeta_{i})  }{\zeta_i - \frac{1}{\mi r}} +
		\frac{ C_n (t, -\zeta^*_i) \psi^{+}_{n,1}(t,-\zeta^*_{i}) }{-\zeta^*_i - \frac{1}{\mi r}}
		\Big) \nonumber \\
		&+
		\frac{1}{2 \pi \mi}   \int_{w\in \Sigma} \frac{ \lambda^{2n}(w) }{w -  \frac{1}{\mi r}}   \bar{\rho}(w)     \psi^{+}_{n,1}(t,w)dw. \label{hlj3.22b}
	\end{align}
\end{subequations}
From the first row in Eq.~\eref{hlj3.22a} and the second row in Eq.~\eref{hlj3.22b}, we have the reconstruction formulas:
\begin{subequations}
	\label{hlj3.23}
	\begin{align}
		Q_{n}(t)=
		&Q_+ +
		\Delta_n(t)
		\displaystyle\sum_{i = 1}^{n_1}
		\Big(
		\frac{\mi r \bar{C}_n (t, \hat{\zeta}_i) \psi^{+}_{n,12}(t,\hat{\zeta}_{i})     }{(\hat{\zeta}_i -\mi r)\hat{\zeta}_i} -
		\frac{\mi r \bar{C}_n (t, -\hat{\zeta}_i^*)  \psi^{+}_{n,12}(t,-\hat{\zeta}^*_{i}) }{(-\hat{\zeta}^*_i -\mi r)\hat{\zeta}^*_i }
		\Big)\nonumber \\
		&
		+  \frac{\Delta_n(t)}{2 \pi \mi}  \int_{w\in \Sigma} \frac{\mi r \lambda^{-2n}(w) }{w (w -\mi r)} \rho(w)  \psi^{+}_{n,12}(t,w)dw,\\
		Q^*_{-n}(t)  =
		&Q^*_{-}
		-\sigma
		\Delta_n(t)
		\displaystyle\sum_{i = 1}^{n_1}
		\Big(
		\frac{ C_n (t, \zeta_i) \psi^{+}_{n,21}(t,\zeta_{i})  }{\zeta_i - \frac{1}{\mi r}} +
		\frac{ C_n (t, -\zeta^*_i) \psi^{+}_{n,21}(t,-\zeta^*_{i}) }{-\zeta^*_i - \frac{1}{\mi r}}
		\Big)\nonumber  \\
		&
		-
		\frac{\sigma \Delta_n(t)}{2 \pi \mi}   \int_{w\in \Sigma} \frac{ \lambda^{2n}(w) }{w -  \frac{1}{\mi r}}   \bar{\rho}(w)     \psi^{+}_{n,21}(t,w)dw.
	\end{align}
\end{subequations}

\section{$N$-soliton solutions in the reflectionless case}
\label{sec4}
In this section, for the reflectionless case, we derive the explicit determinant form of $N$-soliton solutions for Eq.~\eref{NNLS1a} under NZBCs, and then study the dynamical behavior of the obtained soliton solutions. For convenience,  we use the symbol $\Gamma$ to denote the circle  $|z| = 1$.

\subsection{Determinant representation of $N$-soliton solutions}
Under the case of reflectionless potentials ($\rho(\zeta) =  \bar{\rho}(\zeta) = 0$ for $\zeta \in \Sigma$),
the first rows in Eqs.~\eref{hlj139},~\eref{hlj139a} and~\eref{hlj3.22b} constitute a  linear system of $4n_1+1$ equations for
$\psi^{+}_{n,11}(t,\zeta_{i})$,$\psi^{+}_{n,11}(t,-\zeta^*_{i})$, $\psi^{+}_{n,12}(t,\hat{\zeta}_{i})$, $\psi^{+}_{n,12}(t,-\hat{\zeta}^*_{i})$ $(i= 1, \dots, n_1)$ and
$\Delta_n(t)$. By solving those equations, we can obtain the following result:

\begin{theorem}
	\label{Thm5.1}
	Suppose that $\{\zeta_i,-\zeta_i^*\}^{n_1}_{i=1}$ and $\{\hat{\zeta}_i,-\hat{\zeta}_i^*\}^{n_1}_{i=1}$ are the simple poles of scattering coefficients $s_{11}(\zeta)$ and $s_{22}(\zeta)$, respectively, and define that
	\begin{equation}
		\begin{aligned}
			&\eta = \{\eta_i\}_{i = 1}^{2n_1} = \{\zeta_1, \dots, \zeta_{n_1}, -\zeta^*_1, \dots, -\zeta^*_{n_1}\},\\
			&\hat{\eta} = \{\hat{\eta}_i\}_{i=1}^{2n_1}  = \{\hat{\zeta}_1, \dots, \hat{\zeta}_{n_1}, -\hat{\zeta}^*_1, \dots, -\hat{\zeta}^*_{n_1}\}.
		\end{aligned}
	\end{equation}
	Then, the $N$-soliton solutions $(N:=2 n_1)$ of Eq.~\eref{NNLS1a} under NZBCs~\eref{NZBC} in the reflectionless case can be represented by
	\begin{equation}
		\label{Nfold2}
		\begin{aligned}
			&Q_n(t)
			=Q_{+}  + \frac{\det(\mathbf{G}^{aug})}{\det(\mathbf{G})},
		\end{aligned}
	\end{equation}
	with
	\begin{equation}
		\begin{aligned}
			&\mathbf{G} = \begin{pmatrix}
				\mathbf{G}^{(1)} & -\mathbf{I}                 & 0 \\
				-\mathbf{I}                & \mathbf{G}^{(2)}      & G^{(3)} \\
				0                & G^{(4)}   & 1 \\
			\end{pmatrix}, \quad
		   \mathbf{G}^{aug} = \begin{pmatrix}
		   	0 & H  \\
		   	K & \mathbf{G}  \\
		   \end{pmatrix},  \\
	       &\mathbf{G}^{(k)}= \Big(G^{(k)}_{hl}\Big)_{1\leq h,l \leq 2n_1}, \quad k=1,2,  \\
	       &G^{(3)} = \Big(G^{(3)}_{1},\dots,G^{(3)}_{ 2n_1}\Big)^{T},  \quad
	       G^{(4)} = \Big(G^{(4)}_{1},\dots,G^{(4)}_{ 2n_1}\Big), \\
	       &H = (H_{1},\dots,H_{4n_1+1}), \quad
	       K = (K_{1},\dots,K_{ 4n_1+1})^{T},
		\end{aligned}
	\end{equation}
	where
	\begin{equation}
		\begin{aligned}
			&G^{(1)}_{hl} =
			\begin{aligned}
				& \frac{(\eta_{h} - \mi r) \bar{C}_n(t, \hat{\eta}_l)    }{(\hat{\eta}_l - \mi r)(\eta_{h} - \hat{\eta}_l)}, \quad
				h, l \in  [1,2n_1], \\
			\end{aligned}
			 \\ \nonumber
			&G^{(2)}_{hl} =
			\begin{aligned}
				& \frac{(\hat{\eta}_{h} - \frac{1}{\mi r}) C_n(t, \eta_l)   }{(\eta_l - \frac{1}{\mi r})(\hat{\eta}_{h} - \eta_l)}, \quad
				h, l \in  [1,2n_1], \\
			\end{aligned}
			 \\ \nonumber
			&G^{(3)}_{h} =
			\begin{aligned}
				&  \mi r - \frac{1}{\hat{\eta}_{h}}, \quad \quad \quad\quad\quad \, \,\,
				h \in [1,2n_1], \\
			\end{aligned}
			 \\ \nonumber
			&G^{(4)}_{l} =
			\begin{aligned}
				& \frac{ C_n(t, \eta_l)   }{\mi r\eta_l - 1}, \quad \quad\quad\quad \, \,\,
				l \in  [1,2n_1], \\
			\end{aligned}
			 \nonumber
		\end{aligned}
	\end{equation}
	and
	\begin{equation}
		\begin{aligned}
			\quad H_{l} &=
			\left\{
			\begin{aligned}
				&-\frac{\mi r \bar{C}_n(t, \hat{\eta}_l)  }{\hat{\eta}_l(\hat{\eta}_l -\mi r)}, \quad
				l \in[1, 2n_1], \\
				&0, \quad \quad\quad\quad\quad\quad \,\,\,\,
				l  \in[2n_1+1, 4n_1+1], \\
			\end{aligned}
			\right. \\ \nonumber
			K_{h} &=
			\left\{
			\begin{aligned}
				&-Q_{+}, \,\,
				h \in [1,2n_1], \\
				&0, \,\,\quad\quad\,
				h \in [2n_1+1,4n_1], \\
				&1, \,\,\quad\quad\,
				h = 4n_1+1. \\
			\end{aligned}
			\right. \,\, \nonumber
		\end{aligned}
	\end{equation}
\end{theorem}

\begin{remark}
	Considering that there may exist pure imaginary eigenvalues, we denote them as $\zeta_i = \mi \kappa_i$ ($i=1,\dots,m_1$) with $0<\kappa_i<r$ and $0 \leq  m_1 \leq n_1$. In the reflectionless case, the trace formulas~\eref{hlj3.7} reduce to the following expressions:
	\begin{equation}
		\begin{aligned}
			&s_{11}(\zeta)  = \displaystyle\prod_{i = 1}^{l_1} \frac{ \zeta - \hat{\zeta}_i  }{\zeta- \zeta_i }   \frac{ \zeta + \hat{\zeta}^*_i }{\zeta + \zeta^*_i }
			\displaystyle\prod_{i = 1}^{m_1} \frac{ \zeta - \mi \frac{ r \kappa_i + 1}{\kappa_i - r} }{\zeta-  \mi \kappa_i },  \quad
			s_{22}(\zeta)  = c_{\infty}  \displaystyle\prod_{i = 1}^{l_1} \frac{ \zeta - \zeta_i  }{\zeta- \hat{\zeta}_i }   \frac{ \zeta + \zeta^*_i }{\zeta + \hat{\zeta}^*_i }
			\displaystyle\prod_{i = 1}^{m_1} \frac{ \zeta - \mi \kappa_i  }{\zeta- \mi \frac{ r \kappa_i + 1}{\kappa_i - r} },
		\end{aligned}
	\end{equation}
with $l_1+m_1 = n_1$. So, solution~\eref{Nfold2} can represent the general $N$-soliton solutions with $N:=2l_1+m_1$.
In subject to the asymptotic limits of $s_{11}(\zeta)$ and $s_{22}(\zeta)$ as $\zeta \to \frac{1}{\mi r}, 0, \mi r$ in Eq.~\eref{hlj2.51}, there are the following constraints:
	\begin{subequations}
		\label{xcx111}
		\begin{align}
			&\displaystyle\prod_{i = 1}^{l_1} \Big|\frac{  \frac{1}{\mi r} - \hat{\zeta}_i  }{\frac{1}{\mi r}- \zeta_i }\Big|^2
			\displaystyle\prod_{i = 1}^{m_1} \frac{ \frac{1}{\mi r} - \mi \frac{ r \kappa_i + 1}{\kappa_i - r} }{\frac{1}{\mi r}-  \mi \kappa_i } = e^{\mi(\theta_- - \theta_{+})}, \\
			&c_{\infty}  \displaystyle\prod_{i = 1}^{l_1} \Big| \frac{  \zeta_i  }{ \hat{\zeta}_i } \Big|^2
			\displaystyle\prod_{i = 1}^{m_1} \frac{ \kappa_i  (\kappa_i - r) }{r \kappa_i + 1 } = 1, \label{xcx111b}\\
			&c_{\infty}  \displaystyle\prod_{i = 1}^{l_1} \Big|\frac{ \mi r - \zeta_i  }{\mi r- \hat{\zeta}_i} \Big|^2
			\displaystyle\prod_{i = 1}^{m_1} \frac{\mi r - \mi \frac{ r \kappa_i + 1}{\kappa_i - r}  }{\mi r- \hat{\tilde{\zeta}}_i } = e^{\mi(\theta_- - \theta_{+})}. \label{xcx111c}
		\end{align}
	\end{subequations}
By eliminating $c_{\infty}$ from Eqs.~\eref{xcx111b} and~\eref{xcx111c}, we have
\begin{subequations}
	\label{xcx1111}
	\begin{align}
		&\displaystyle\prod_{i = 1}^{l_1} \Big|\frac{  \frac{1}{\mi r} - \hat{\zeta}_i  }{\frac{1}{\mi r}- \zeta_i }\Big|^2
		\displaystyle\prod_{i = 1}^{m_1} \frac{(\kappa_i - r)\big(1+\kappa_i r\big)}{\kappa_i Q^2_0} = e^{\mi(\theta_- - \theta_{+})}, \label{xcx1111a}\\
		&\displaystyle\prod_{i = 1}^{l_1} \Big|\frac{ \mi r - \zeta_i  }{\mi r- \hat{\zeta}_i} \Big|^2
		\displaystyle\prod_{i = 1}^{m_1} \frac{\kappa_i Q^2_0}{(\kappa_i - r)\big(1+\kappa_i r\big)} = e^{\mi(\theta_- - \theta_{+})},\label{xcx1111b}
	\end{align}
\end{subequations}
which means that $m_1$ must be odd if $Q_{+}=Q_{-}$, and even if $Q_{+}=-Q_{-}$.
\end{remark}

\subsection{Soliton solutions with large NZBCs}
Based on constraints~\eref{xcx1111a} and~\eref{xcx1111b}, we can properly select the integers $l_1, m_1$ with given $\sigma$ and $\Delta \theta$.
For the case $\sigma = 1$, the eigenvalues appear in pairs on the $\Gamma$ or are symmetrically distributed in pairs about the $\Gamma$.
The resulting solutions always exhibit singularities in the continuous limit, so we will not discuss them in detail.
Differently, in the case of $\sigma=-1$, there is at least one pure imaginary eigenvalue located on $\Gamma$, and the solutions can exhibit the soliton behavior with a wide range of parameter regimes.
In the following, we study the dynamical behavior of soliton solutions for three different cases with  $N \leq 3$.

\begin{figure}[H]
	\centering
	\subfigure[]{
		\label{f1a}
		\includegraphics[width=1.9in]{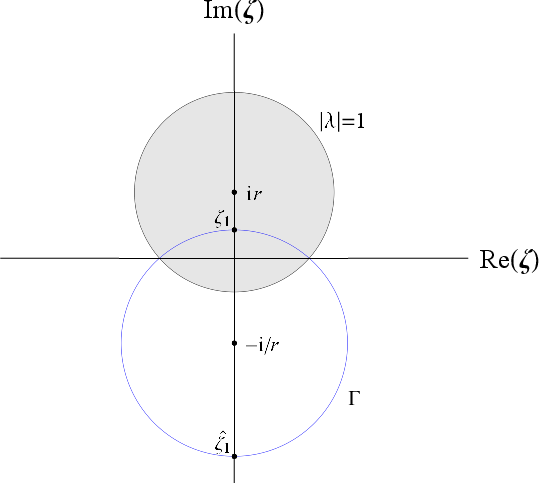}}\hfill
	\subfigure[]{
		\label{f1b}
		\includegraphics[width=1.9in]{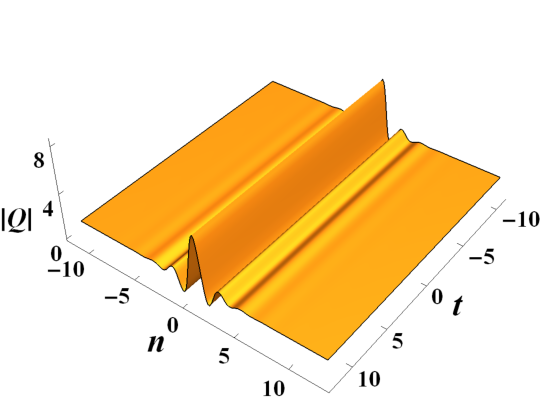}}\hfill
	\subfigure[]{
		\label{f1c}
		\includegraphics[width=1.9in]{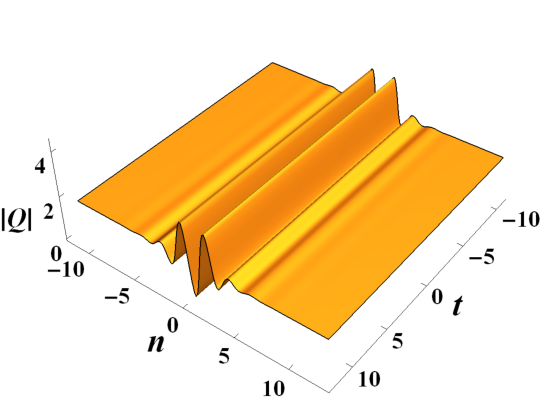}}\hfill
	\caption{\small
		(a) Distribution of discrete eigenvalues $\{\zeta_{1}, \hat{\zeta}_{1}\}$;
		(b) An oscillating anti-dark soliton with $Q_+ = 2 e^{\frac{\pi}{3} \mi}$, $\zeta_{1} =  \frac{\sqrt{3} \mi }{3} $, $\hat{\zeta}_{1} = -\sqrt{3} \mi$ and $\theta_{1} = 2$.
		(c) An oscillating dark soliton with $Q_+ = 2 e^{\frac{\pi}{3}\mi}$, $\zeta_{1} =  \frac{\sqrt{3} \mi }{3} $, $\hat{\zeta}_{1} = \sqrt{3} \mi$ and $\theta_{1} = 0$.
		\label{f1} }
\end{figure}

\textbf{Case 4.1} $l_1 = 0$ and $m_1 = 1$. In this case,  the sole pure imaginary eigenvalue $\zeta_{1} = \frac{1}{\mi r} + \frac{\mi Q_0}{ r}$ is located on the circle $\Gamma$ (see Fig.~\ref{f1a}). According to Eqs.~\eref{hlj3.14}, the norming constants $\overline{s}_{12}(\zeta_{1})$ and $\overline{s}_{21}(\hat{\zeta}_1)$ meet the following constraints:
\begin{equation}
	\begin{aligned}
		|\overline{s}_{12}(\zeta_{1})|^2 = |\overline{s}_{21}(\hat{\zeta}_1)|^2  = (Q_0 + 1)^6, \quad \overline{s}_{12}(\zeta_{1}) = \overline{s}_{21}(\hat{\zeta}_1) e^{-2 \mi \theta_{+}}.
	\end{aligned}
\end{equation}
Then, we can present the following solution:
\begin{equation}
	\label{4.8}
	\begin{aligned}[b]
		Q_n(t)
		&=  Q_+ \frac{ \left(e^{2 \mi \theta _+} r^{4 n}-e^{2 \mi b} \left(Q_0-1\right){}^{4 n}\right)-2 \mi e^{\mi n \pi }  e^{\mi \left(b+ \theta _+ \right)} r^{2 n} \left(Q_0-1\right)^{2 n}}
		{e^{2 \mi \theta _+} r^{4 n} + e^{2 \mi b} \left(Q_0-1\right){}^{4 n}} \noindent \\
		&= Q_+ \big[  \tanh (\eta_n) - \mi (-1)^n \sech (\eta_n) \big] ,
	\end{aligned}
\end{equation}
where $\eta_n = 2 n \ln \Big(\frac{r}{Q_0-1}\Big) + \mi \theta_{+} - \mi \theta_{1}$ with $\theta_{1} = \arg(\overline{s}_{21}(\hat{\zeta}_1))$.
It should be noted that the term $(-1)^n$ in Eq.~\eref{4.8} originates from the calculation of $\lambda^{2n}(\zeta_{1}) = \left(\frac{1-Q_0}{1+Q_0}\right)^n$ with $\frac{1-Q_0}{1+Q_0}<0$ in Eq.~\eref{hlj3.14}, which is sharp contrast with $\frac{1-Q_0}{1+Q_0}>0$ for small NZBCs~\cite{YHLiu}.
As a result, the heteroclinic solution~\eref{4.8} shows localized structure along with $n$-dependent periodical oscillation.
Through calculation, the square moduli $|Q_n(t)|^2$ can be expressed as
\begin{equation}
	\label{4.9}
	\begin{aligned}
		&|Q_n(t)|^2 = Q_0^2 \Bigg(1 - \frac{2\sin(\eta_{n,I})}{\sin(\eta_{n,I}) + \cosh( 2 \eta_{n,R} )}\Bigg),
	\end{aligned}
\end{equation}
with the subscripts $R$ and $I$, respectively, denoting the real and imaginary parts of $\eta_n$, which implies that this solution is non-singular if and only if $\theta_{+} - \theta_{1} \neq -\frac{\pi}{2} + 2 m \pi$ ($m \in \mathbb{Z}$).
Particularly at $n=0$, we have
\begin{equation}
	\label{4.10}
	\begin{aligned}
		|Q_0(t)|^2 = Q^2_0 \Bigg(\frac{1 - \sin(\theta_{+} -  \theta_{1})}{1 +  \sin(\theta_{+} -  \theta_{1}) }\Bigg),
	\end{aligned}
\end{equation}
which corresponds to the main valley for $2 m \pi <\theta_{+} - \theta_{1}<(2m+1)\pi$ ($m \in \mathbb{Z}$), or the main peak if $(2m-1)\pi<\theta_{+} - \theta_{1}<2 m \pi$ and $\theta_{+} - \theta_{1} \neq -\frac{\pi}{2} + 2 m \pi$ ($m \in \mathbb{Z}$). Therefore, solution~\eref{4.10} can display the oscillating anti-dark and dark solitons, as depicted in Figs.~\ref{f1b} and~\ref{f1c}.
In addition, if $ \theta_{+} = \theta_{1} + m \pi$ ($m \in \mathbb{Z}$), solution~\eref{4.8} will degenerate to a plane wave.
%
%

\begin{figure}[H]
	\centering
	\subfigure[]{
		\label{f2a}
		\includegraphics[width=1.9in]{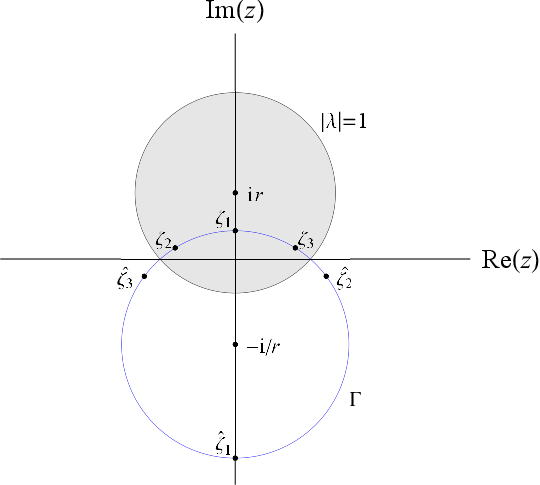}}\hfill
	\subfigure[]{
		\label{f2b}
		\includegraphics[width=1.9in]{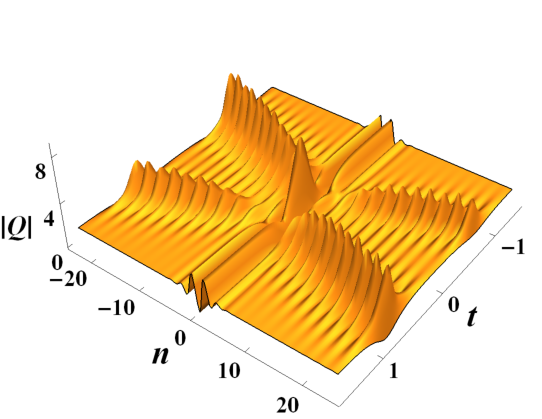}}\hfill
	\subfigure[]{
		\label{f2c}
		\includegraphics[width=1.9in]{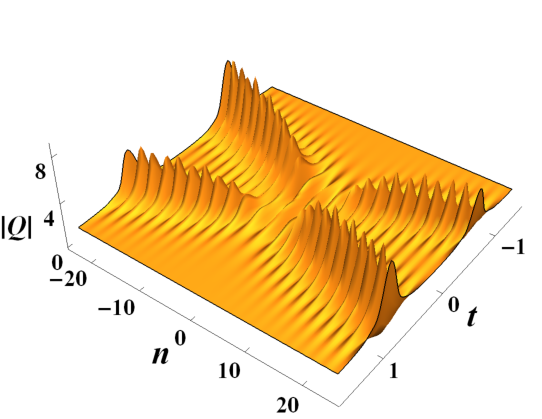}}\hfill
	\caption{\small
		(a) Distribution of discrete eigenvalues $\{\zeta_{1}, \hat{\zeta}_{1}\}$, $\{\zeta_{2}, \hat{\zeta}_{2}\}$ and $\{\zeta_{3}, \hat{\zeta}_{3}\}$;
		(b) An interaction among three oscillating solitons with $Q_+ = 2 e^{\mi}$,
		$\zeta_{1} =  \frac{\sqrt{3} \mi }{3}$, $\zeta_{2} =  \frac{8 \sqrt{3} \mi }{15} + \frac{ \sqrt{3} \mi }{15}$, $\zeta_{3} = - \frac{8 \sqrt{3} \mi }{15} + \frac{ \sqrt{3} \mi }{15}$,
		$\hat{\zeta}_{1} = -\sqrt{3} \mi$, $\hat{\zeta}_{2} = - \frac{8 \sqrt{3} \mi }{13} - \frac{ \sqrt{3} \mi }{13}$, $\hat{\zeta}_{3} =  \frac{8 \sqrt{3} \mi }{13} - \frac{ \sqrt{3} \mi }{13}$,
		$\theta_{1} = 3$, and $\theta_{2} =\theta_{3} = 2$;
		(c) An interaction between two oscillating solitons with $Q_+ = 2 e^{\frac{\pi}{3} \mi }$,
		$\zeta_{1} =  \frac{\sqrt{3} \mi }{3}$, $\zeta_{2} =  \frac{8 \sqrt{3} \mi }{15} + \frac{ \sqrt{3} \mi }{15}$, $\zeta_{3} = - \frac{8 \sqrt{3} \mi }{15} + \frac{ \sqrt{3} \mi }{15}$,
		$\hat{\zeta}_{1} = -\sqrt{3} \mi$, $\hat{\zeta}_{2} = - \frac{8 \sqrt{3} \mi }{13} - \frac{ \sqrt{3} \mi }{13}$, $\hat{\zeta}_{3} =  \frac{8 \sqrt{3} \mi }{13} - \frac{ \sqrt{3} \mi }{13}$,
		$\theta_{1} = \frac{\pi}{3}$, and $\theta_{2} = \theta_{3} = -1$.
		\label{f2} }
\end{figure}

\textbf{Case 4.2} $l_1 = 2$ and $m_1 = 1$.
There are three eigenvalues on the circle $\Gamma$, in which  one pure imaginary eigenvalue $\zeta_1 = \frac{1}{\mi r} + \frac{\mi Q_0}{ r}$ and
two complex eigenvalues $\zeta_{2} = \frac{1}{\mi r} + \frac{Q_0}{ r} e^{\mi \varphi }, \zeta_{3} =  \frac{1}{\mi r} - \frac{Q_0}{ r} e^{-\mi \varphi}$ with $\arctan(r)<\varphi < \frac{\pi}{2}$ (see Fig.~\ref{f2a}).
Then, based on Eq.~\eref{hlj3.14}, we obtain the following constraints on the norming constants:
\begin{equation}
	\begin{aligned}
		&\overline{s}_{21}(\hat{\zeta}_1)  = \frac{ \mi r^5}{(Q_0-1)^3 |Q^2_0 - 2 \sin(\varphi) Q_0 -1| } e^{\mi \theta_{1}}, \quad
		\overline{s}_{12}(\zeta_{1}) = \overline{s}_{21}(\hat{\zeta}_1) e^{-2 \mi \theta_{+}}, \\
		&\overline{s}_{21}(\hat{\zeta}_2)  = \frac{  r^5}{(Q_0-1) (Q^2_0 - 2 \sin(\varphi) Q_0 -1)^2 } e^{\mi \theta_{2}}, \quad \overline{s}_{12}(\zeta_{2}) = \overline{s}_{21}(\hat{\zeta}_2) e^{-2 \mi \theta_{+}}, \\
		&\overline{s}_{21}(\hat{\zeta}_3)  = \frac{  r^5}{(Q_0-1) (Q^2_0 - 2 \sin(\varphi) Q_0 -1)^2 } e^{\mi \theta_{3}}, \quad \overline{s}_{12}(\zeta_{3}) = \overline{s}_{21}(\hat{\zeta}_3) e^{-2 \mi \theta_{+}},
	\end{aligned}
\end{equation}
where $\theta_{1} = \arg(\overline{s}_{21}(\zeta_1))$, $\theta_{2} = \arg(\overline{s}_{21}(\hat{\zeta}_2)) $ and $\theta_{3}  = \arg(\overline{s}_{21}(\hat{\zeta}_3)) $.

Note that the ($n,t$)-dependent parts  in Eq.~\eref{hlj3.14} can be written in the exponential form:
\begin{equation}
	\label{xcx1021}
	\begin{aligned}
		&\lambda(\zeta_{1})^{2 n } e^{\Omega_{11}(\zeta_{1}) t - \Omega_{22}(\zeta_{1}) t} =
		 e^{2 n \ln\Big(\frac{Q_0-1}{r}\Big) + \mi n \pi }, \\
		&\lambda(\zeta_{2})^{2 n } e^{\Omega_{11}(\zeta_{2}) t - \Omega_{22}(\zeta_{2}) t}
			= e^{n \ln\Big(\frac{Q^2_0 - 2 \sin(\varphi) Q_0 -1}{r^2}\Big) + \mi n \pi - \frac{4 \cos(\varphi) (Q_0 - \sin(\varphi))}{Q^2_0 - 2 \sin(\varphi) Q_0 -1} t},\\
		&\lambda(\zeta_{3})^{2 n } e^{\Omega_{11}(\zeta_{3}) t - \Omega_{22}(\zeta_{3}) t}
			= e^{n \ln\Big(\frac{Q^2_0 - 2 \sin(\varphi) Q_0 -1}{r^2}\Big) + \mi n \pi + \frac{4 \cos(\varphi) (Q_0 - \sin(\varphi))}{Q^2_0 - 2 \sin(\varphi) Q_0 -1} t}.
	\end{aligned}
\end{equation}
Then, by respectively assuming that
\begin{equation}
	\begin{aligned}
		&\Theta_1: = 2 n \ln\Big(\frac{Q_0-1}{r}\Big)= O(1), \\
		&\Theta_2: = n \ln\Big(\frac{Q^2_0 - 2 \sin(\varphi) Q_0 -1}{r^2}\Big) - \frac{4 \cos(\varphi) (Q_0 - \sin(\varphi))}{Q^2_0 - 2 \sin(\varphi) Q_0 -1} t = O(1),\\
		&\Theta_3: = n \ln\Big(\frac{Q^2_0 - 2 \sin(\varphi) Q_0 -1}{r^2}\Big) + \frac{4 \cos(\varphi) (Q_0 - \sin(\varphi))}{Q^2_0 - 2 \sin(\varphi) Q_0 -1} t = O(1),
	\end{aligned}
\end{equation}
we obtain that there are three asymptotic solitons as $t \to \pm \infty$:
\begin{equation}
	\label{asy}
	\begin{aligned}
		&q^{\pm}_{1} = - Q_+ \frac{e^{\pm \mi \varphi } \mp \mi Q_0}{e^{\mp \mi \varphi } \pm \mi Q_0} \Big[ \tanh(\Theta_1  + \mi  (\theta_{1}-\theta_{+}) )
		+ (-1)^{n+\frac{3}{2}}  \sech(\Theta_1 + \mi  (\theta_{1}-\theta_{+}) )  \Big], \\
		&q^{\pm}_{2} = Q_+ \frac{\sin(\varphi) - Q_0}{\mp \mi e^{ \mp \mi \varphi} + Q_0} \Big[ \tanh(\Theta_2  + \ln( \tau_1^{\pm}  ) )
		+ (-1)^{n+\frac{3}{2}}  \sech(\Theta_2 + \ln(\tau_1^{\pm} ) )
		\Big] \pm \frac{ Q_+ \cos(\varphi) }{ e^{\mp \mi \varphi} \pm \mi Q_0}, \\
		&q^{\pm}_{3} = Q_+ \frac{\sin(\varphi) - Q_0}{\mp \mi e^{ \mp \mi \varphi} + Q_0} \Big[ \tanh(\Theta_3  + \ln( \tau_1^{\mp}  ) )
		+ (-1)^{n+\frac{3}{2}}  \sech(\Theta_3 + \ln(\tau_1^{\mp}  ) )
		\Big] \mp \frac{ Q_+ \cos(\varphi) }{ e^{\mp \mi \varphi} \pm \mi Q_0},
	\end{aligned}
\end{equation}
with
\begin{equation}
	\begin{aligned}
		&\tau_1^{\pm} =  - \mi e^{\mi(\theta_{2} - \theta_{+})} \Bigg[\frac{ 2   e^{\mi \varphi } (\sin(\varphi) Q_0 - 1)}{(1+Q_0) (e^{\mi \varphi}-\mi)^2 }\Bigg]^{\pm 1}.
	\end{aligned}
\end{equation}
Thus, the solution consists of a static oscillating soliton and two propagating solitons with velocities $v^{\pm}_2 = -v^{\pm}_3 =  \frac{4 \cos(\varphi) (Q_0 - \sin(\varphi))}{(Q^2_0 - 2 \sin(\varphi) Q_0 -1) \ln\big(\frac{Q^2_0 - 2 \sin(\varphi) Q_0 -1}{r^2}\big) }$, as seen in Fig.~\ref{f2b}.

Moreover, the square moduli of asymptotic solitons can be calculated as follows:
\begin{subequations}
	\begin{align}
		&|q^{\pm}_{1}|^2 = Q^2_0 \Bigg( \frac{1 - \sin(\theta_{+} -  \theta_{1})}{1 +  \sin(\theta_{+} -  \theta_{1}) }\Bigg),\\
		&|q^{\pm}_{2}|^2 = Q_0^2 - Q_0^2 \Bigg(\frac{2(Q_0 - \sin(\varphi) )(\cos(\theta_2 - \theta_{+} +\varphi)+Q_0 \sin(\theta_2 - \theta_{+}))}{(\sin(\theta_2 - \theta_{+}) + (-1)^{n}) (Q_0^2 - 2\sin(\varphi) Q_0 + 1 )}\Bigg), \label{4.15b}\\
		&|q^{\pm}_{3}|^2 = Q_0^2 + Q_0^2 \Bigg(\frac{2(Q_0 - \sin(\varphi) )(\cos(\theta_2 - \theta_{+} +\varphi)-Q_0 \sin(\theta_2 - \theta_{+}))}{(\sin(\theta_2 - \theta_{+}) + (-1)^{n}) (Q_0^2 - 2\sin(\varphi) Q_0 + 1 )}\Bigg).
		\label{4.15c}
	\end{align}
\end{subequations}
Apparently, $q^{\pm}_{1}$ has no singularity if and only if $\theta_{+} - \theta_{1} \neq -\frac{\pi}{2} + 2 m \pi$ ($m \in \mathbb{Z}$), and both $q^{\pm}_{2}$ and $q^{\pm}_{3}$ have no singularity if and only if $\theta_{2} - \theta_{+} \neq \frac{\pi}{2} + m \pi$ ($m \in \mathbb{Z}$).
Similarly to Eq.~\eref{4.8}, $q^{\pm}_{1}$ can also display the oscillating dark soliton for $2 m \pi <\theta_{+} - \theta_{1}<(2m+1)\pi$ ($m \in \mathbb{Z}$) and anti-dark soliton $(2m-1)\pi<\theta_{+} - \theta_{1}<2 m \pi$ and $\theta_{+} - \theta_{1} \neq -\frac{\pi}{2} + 2 m \pi$ ($m \in \mathbb{Z}$).
Differently, the amplitudes of asymptotic solitons $q^{\pm}_{2}$ and $q^{\pm}_{3}$ oscillate between the odd and even numbers of $n$.
When $n$ is even (or odd) and $\theta_{2} - \theta_{+} \neq \frac{\pi}{2} + m \pi$ ($m \in \mathbb{Z}$),
$q^{\pm}_{\rm{2}}$ represents the anti-dark (or dark) soliton for $\frac{\cos(\theta_2 - \theta_{+} +\varphi)}{\sin(\theta_2 - \theta_{+})} < -Q_0$ and dark (or anti-dark) soliton for $\frac{\cos(\theta_2 - \theta_{+} +\varphi)}{\sin(\theta_2 - \theta_{+})} > -Q_0$,
whereas
$q^{\pm}_{\rm{3}}$ represents the anti-dark (or dark) soliton for $\frac{\cos(\theta_2 - \theta_{+} +\varphi)}{\sin(\theta_2 - \theta_{+})} > Q_0$ and dark (or anti-dark) soliton for $\frac{\cos(\theta_2 - \theta_{+} +\varphi)}{\sin(\theta_2 - \theta_{+})} < Q_0$.
Therefore, both $q^{\pm}_{2}$ and $q^{\pm}_{3}$ exhibit an alternating profiles of dark and anti-dark soliton as $n$ changes.
In addition, if $ \theta_{1} - \theta_{+} =   m \pi$ (or $ |\frac{\cos(\theta_2 - \theta_{+} +\varphi)}{\sin(\theta_2 - \theta_{+})}| = Q_0$) ($m \in \mathbb{Z}$), the asymptotic soliton $q^{\pm}_{1}$, $q^{\pm}_{2}$ or $q^{\pm}_{3}$ degenerates into the plane wave and the corresponding solution just shows an interaction between two oscillating solitons (seen in Fig.~\ref{f2c}).

\begin{figure}[H]
	\centering
	\subfigure[]{
		\label{af3a}
		\includegraphics[width=1.9in]{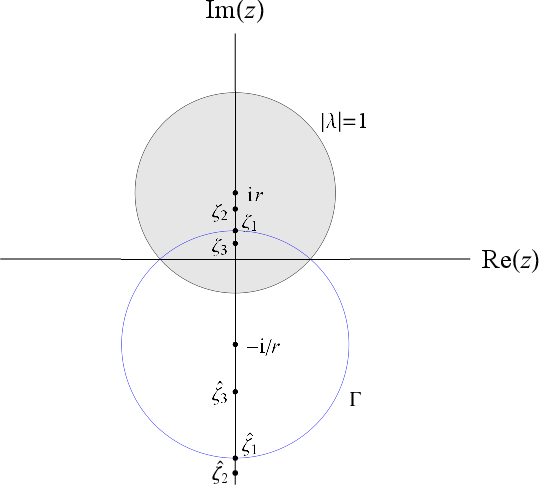}}\hfill
	\subfigure[]{
		\label{af3b}
		\includegraphics[width=1.9in]{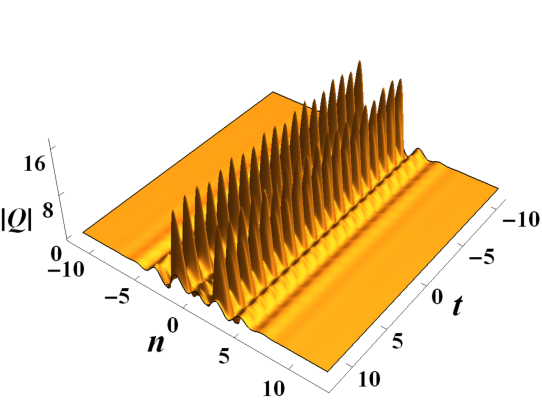}}\hfill
	\subfigure[]{
		\label{af3c}
		\includegraphics[width=1.9in]{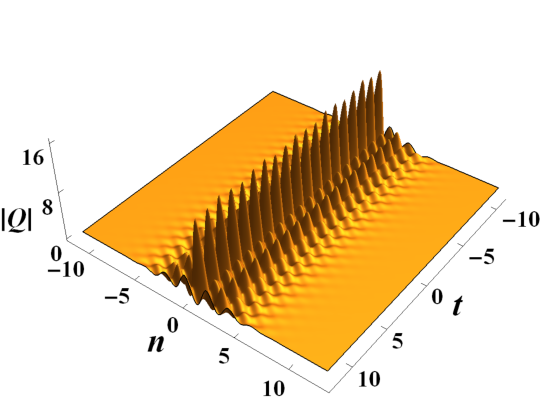}}\hfill
	\caption{\small
		(a) Distribution of discrete eigenvalues $\{\zeta_{1}, \hat{\zeta}_{1}\}$, $\{\zeta_{2}, \hat{\zeta}_{2}\}$ and $\{\zeta_{3}, \hat{\zeta}_{3}\}$;
		(b) Superposition of one oscillating dark soliton and one breather with $Q_+ = 2 e^{\frac{\pi}{3} \mi }$,
		$\zeta_{1} =  \frac{\sqrt{3} \mi }{3}$, $\zeta_{2} =  \frac{\sqrt{3} \mi }{6}$, $\zeta_{3} =  \frac{5\sqrt{3} \mi }{9}$,
		$\hat{\zeta}_{1} = -\sqrt{3} \mi$, $\hat{\zeta}_{2} = - \frac{3\sqrt{3} \mi }{5}$, $\hat{\zeta}_{3} = -2 \sqrt{3} \mi$,
		$\theta_{1} = 0$, $\theta_{2} =  \frac{\pi}{4}$ and $\theta_{3} = \frac{\pi}{4}$;
		(c) One breather with $Q_+ = 2 e^{\frac{\pi}{4} \mi}$,
		$\zeta_{1} =  \frac{\sqrt{3} \mi }{3}$, $\zeta_{2} =  \frac{\sqrt{3} \mi }{6}$, $\zeta_{3} =  \frac{5\sqrt{3} \mi }{9}$,
		$\hat{\zeta}_{1} = -\sqrt{3} \mi$, $\hat{\zeta}_{2} = - \frac{3\sqrt{3} \mi }{5}$, $\hat{\zeta}_{3} = -2 \sqrt{3} \mi$,
		$\theta_{1} = \frac{\pi}{4}$, $\theta_{2} =  2$ and $\theta_{3} = 2$.
		\label{af3} }
\end{figure}

\textbf{Case 4.3} $l_1 = 0$ and $m_1 = 3$. There are three pure imaginary eigenvalues $\zeta_{1} = \frac{1}{\mi r} + \frac{\mi Q_0}{ r}$, $\zeta_{2} = \frac{1}{\mi r} + \frac{\mi \kappa}{ r}$ and $\zeta_{3} = \frac{1}{\mi r} + \frac{\mi Q^2_0}{ \kappa r}$ with $1<\kappa < Q^2_0$, in which $\zeta_{1}$ is on the circle $\Gamma$, $\zeta_{2}$ and $\zeta_{3}$ are symmetrically distributed with respect to $\Gamma$ (see Fig.~\ref{af3a}).
In this case, Eq.~\eref{hlj3.14} implies the following constraints on the norming constants $\overline{s}_{12}(\zeta_{i})$ and $\overline{s}_{21}(\hat{\zeta}_i)$ ($i =1,2,3$):
\begin{equation}
	\begin{aligned}
		&\overline{s}_{21}(\hat{\zeta}_1)  = \frac{ \kappa r^5}{(Q_0-1)^3 (Q^2_0 - \kappa) (\kappa-1)} e^{\mi \theta_{1}}, \quad \overline{s}_{12}(\zeta_{1}) = \overline{s}_{21}(\hat{\zeta}_1) e^{-2 \mi \theta_{+}}, \\
		&\overline{s}_{21}(\hat{\zeta}_2)  = \frac{ \kappa^2 r^5}{(Q_0-1) (Q^2_0 - \kappa)^2 (\kappa-1)^2} e^{\mi \theta_{2}}, \quad \overline{s}_{12}(\zeta_{2}) = \overline{s}_{21}(\hat{\zeta}_2) e^{-2 \mi \theta_{+}}, \\
		&\overline{s}_{21}(\hat{\zeta}_3)  = \frac{ \kappa^2 r^5}{(Q_0-1) (Q^2_0 - \kappa)^2 (\kappa-1)^2} e^{\mi \theta_{3}}, \quad \overline{s}_{12}(\zeta_{3}) = \overline{s}_{21}(\hat{\zeta}_3) e^{-2 \mi \theta_{+}},
	\end{aligned}
\end{equation}
where $\theta_{1}= \arg(\overline{s}_{21}(\hat{\zeta}_1))$, $\theta_{2}= \arg(\overline{s}_{21}(\hat{\zeta}_2))$ and $\theta_{3}= \arg(\overline{s}_{21}(\hat{\zeta}_3))$.

Again, the ($n,t$)-dependent parts in Eq.~\eqref{hlj3.14} can be written in the exponential form
\begin{equation}
	\label{xcx921}
	\begin{aligned}
		&\lambda(\zeta_{1})^{2 n } e^{\Omega_{11}(\zeta_{1}) t - \Omega_{22}(\zeta_{1}) t} =  e^{n \ln\big(\frac{(Q_0-1)^2}{r^2}\big) + \mi n \pi }, \\
		&\lambda(\zeta_{2})^{2 n } e^{\Omega_{11}(\zeta_{2}) t - \Omega_{22}(\zeta_{2}) t}
		= e^{n \ln\big( \frac{(\kappa - 1) (\kappa - Q_0^2)}{\kappa (Q_0^2-1)}   \big) + \mi n \pi - \frac{\mi  \left(\kappa^2 - Q_0^2 \right)\left(\kappa^2 - 2 \kappa Q_0^2 +  Q_0^2 \right) }{\kappa (\kappa - 1) (\kappa - Q^2_0)} t },\\
		&\lambda(\zeta_{3})^{2 n } e^{\Omega_{11}(\zeta_{3}) t - \Omega_{22}(\zeta_{3}) t}
		= e^{n \ln\big(  \frac{(\kappa - 1) (\kappa - Q_0^2)}{\kappa (Q_0^2-1)}   \big) + \mi n \pi + \frac{\mi  \left(\kappa^2 - Q_0^2 \right)\left(\kappa^2 - 2 \kappa Q_0^2 +  Q_0^2 \right) }{\kappa (\kappa - 1) (\kappa - Q^2_0)} t}.
	\end{aligned}
\end{equation}
Similar to the previous cases, the imaginary eigenvalue $\zeta_{1}$ corresponds to an oscillating soliton like Eq.~\eref{4.8}, and the eigenvalue pair $(\zeta_{2},\zeta_{3})$ leads to a breather which is localized around $x=0$ and propagates as $t$ evolves with the period $\frac{\kappa (\kappa - 1) (\kappa - Q^2_0) \pi }{\left(\kappa^2 - Q_0^2 \right)\left(\kappa^2 - 2 \kappa Q_0^2 +  Q_0^2 \right)}$. Thus, the resulting solution consists of a breather and an oscillating soliton, both of which have the zero velocities, as seen in Fig.~\ref{af3b}. In particular,  when $\theta_+ =  \theta_{1}$, the oscillating soliton associated to $\zeta_{1}$ vanishes as $t\to \pm \infty$, so we only observe the breathers as shown in Fig.~\ref{af3c}.

\section{Conclusions}
\label{sec5}
In this paper, we extended the IST theory for the discrete PT-symmetric NNLS equation~\eref{NNLSS11} under large NZBCs.
By considering that the data at infinity have constant amplitudes, we studied two cases in which the previous IST formulation does not work for large NZBCs~\cite{YHLiu}.
By introducing a suitable uniformization variable, we provided rigorous proofs of the analyticity, symmetries and asymptotic behaviors of the eigenfunctions and scattering coefficients for the direct problem.
Then, we formulated the RHP for the inverse problem and obtained the integral representation for the solutions with the presence of an arbitrary number of simple zeros.
In sharp contrast to small NZBCs case,  the IST for large NZBCs imposes no restriction on the location of discrete eigenvalues.


On the other hand, we derived the determinant representation of the discrete $N$-soliton solutions and studied the soliton dynamics in the focusing type of equation.
Different from the case of small NZBCs, we revealed two types of novel solitons with large NZBCs: (i) The oscillating soliton, which corresponds to an eigenvalue lying on the unit circle $\Gamma$ and can display the oscillatory dark and anti-dark  profiles along the lattice sites $n$; (ii) The breather, associated with a pair of eigenvalues symmetrically located with respect to $\Gamma$, which features internal periodic oscillation as the time $t$ evolves. Remarkably, the former has not been previously reported for either the PT-symmetric NNLS equation or the AL equation, while the latter does not appear under small NZBCs.
In addition, via the asymptotic analysis method, we demonstrated that the multi-soliton solutions can display a rich variety of interactions, including the collisions among oscillating dark/anti-dark solitons, and the superpositions of oscillating soliton and  breather.

\section*{Data availability statement}

This work does not contain any experimental data. All the mathematical results are in analytic form and are reproducible.

\section*{Funding}
This work was supported by the National Natural Science Foundation of China (Grant No. 12475003),  by the Beijing Natural Science Foundation (Grant No. 1252016), and by the China Scholarship Council (Program No. 202406440126).


\begin{appendix}
\renewcommand{\thesection}{Appendix~\Alph{section}}
\renewcommand{\thesubsection}{\Alph{section}.\arabic{subsection}}

\section{}
\label{appendixA}
\renewcommand{\theequation}{A.\arabic{equation}}
\setcounter{equation}{0}


\subsection{Proof of Eq.~\eref{xcxa21}}
\label{appendixA1a}
For brevity, we omit the $z$-dependence.
First, the inverse of $\mathbf{U}_{n}(t)$ can be given by
\begin{equation}
	\label{hljA.4}
	\begin{aligned}
		\mathbf{U}_{n}^{-1}(t) = \frac{1}{1- \sigma Q_{n}(t) Q^*_{-n}(t) }\begin{pmatrix}
			\frac{1}{z} & -Q_{n}(t) \\
			- \sigma Q^*_{-n}(t) & z \\
		\end{pmatrix}
		=
		\frac{
		\mathbf{Y}_{+}
		(\mi r \mathbf{J})^{-1}
		\mathbf{Y}^{-1}_{+}
		-  \Delta \mathbf{Q}_{n}^{+}(t)
	}{1 - \sigma Q_{n}(t) Q^*_{-n}(t) }.
	\end{aligned}
\end{equation}
Then, left-multiplying both sides of Eq.~\eref{Laxpairjvzhenxingshiaa} by $(\mi r \mathbf{J})^{-1} \mathbf{U}_{n}^{-1}(t) $ and right-multiplying by $\mathbf{Y}^{-1}_{+}$ yields
\begin{equation}
	\begin{aligned}
		&\bm{\mu}_{n}^{+}(t)
		(\mi r \mathbf{J})^{-1} =
		\frac{
	(\mi r \mathbf{J})^{-1}
	-  \mathbf{Y}^{-1}_{+} \Delta  \mathbf{Q}_{n}^{+}(t) \mathbf{Y}_{+}	
	}{1- \sigma Q_{n}(t) Q^*_{-n}(t) }  \bm{\mu}_{n+1}^{+}(t),
	\end{aligned}
\end{equation}
where $\bm{\mu}_{n}^{+}(t)$ has been defined in Eq.~\eref{xcxa19}.
Further, we multiply the above equation by $\displaystyle\prod_{k = n}^{ \infty} (1- \sigma Q_k(t)  Q^*_{-k}(t))$, yielding
\begin{equation}
	\begin{aligned}
		& \bm{\mu}_{n}^{+}(t) \displaystyle\prod_{k = n}^{ \infty} (1- \sigma Q_k(t)  Q^*_{-k}(t))
		(\mi r \mathbf{J})^{-1} =
		\big(
		(\mi r \mathbf{J})^{-1}
		-  \mathbf{Y}^{-1}_{+} \Delta  \mathbf{Q}_{n}^{+}(t) \mathbf{Y}_{+} \big)
		\bm{\mu}_{n+1}^{+} \displaystyle\prod_{k = n+1}^{ \infty} (1- \sigma Q_k(t)  Q^*_{-k}(t)) .
	\end{aligned}
\end{equation}
Finally, by denoting  $\bm{\mu}_{n}^{+}(t) \displaystyle\prod_{k = n}^{ \infty} (1- \sigma Q_k(t)  Q^*_{-k}(t))$ as $\hat{\bm{\mu}}_{n}^{+}(t)$, we obtain Eq.~\eref{xcxa21}.

\subsection{Proof of Theorem~\ref{th1}}
\label{appendixA3}
In this part, for brevity, we omit the $t$-dependence.
Based on Eq.~\eref{xcxa20a}, $\mu^{-}_{n,1}(z)$ can be expressed as the following Neumann series:
\begin{equation}
	\label{xcxa39a}
	\begin{aligned}
		&\mu^{-}_{n,1}(z, \lambda) =\sum_{j=0}^{\infty} w^{(j)}_n(z, \lambda),
	\end{aligned}
\end{equation}
with
\begin{equation}
	\label{xcxa39}
	\begin{aligned}
		&w^{(0)}_n(z, \lambda) =
		\left(
		\begin{array}{ccc}
			1 \\
			0 \\
		\end{array}
		\right), \quad
		w^{(j+1)}_n(z, \lambda) = \displaystyle\sum_{k = -\infty}^{n - 1}  \mathbf{G}_{k}(z, \lambda)	w^{(j)}_{k}(z, \lambda), \\
        &\mathbf{G}_{k}(z, \lambda) =\frac{1}{\mi r \lambda} \begin{pmatrix}
			1  &0  \\
			0 &  \lambda^{-2n+2k+2} \\
		\end{pmatrix}
		\mathbf{Y}^{-1}_{-}(z, \lambda) \Delta \mathbf{Q}_{k}^{-}(t)  \mathbf{Y}_{-}(z, \lambda).
	\end{aligned}
\end{equation}

Introducing the $L^{1}$ vector norm $||\mu^{-}_{n,1}(z, \lambda)||=|\mu^{-}_{n,11}(z, \lambda)|+|\mu^{-}_{n,21}(z, \lambda)|$ and  the corresponding subordinate matrix norm, we then have
\begin{equation}
	\begin{aligned}
		&||\mathbf{Y}_{-}(z, \lambda)|| \leq  Q_0 + |r\lambda| + |z|,\\
		&||\mathbf{Y}^{-1}_{-}(z, \lambda)|| \leq \frac{1}{|\gamma(z) |}  \big(Q_0 + |r\lambda| + |z| \big),\\
		&|| \mathbf{G}_{k}(z, \lambda) || \leq c(z, \lambda) |1 + \lambda^{-2n+2k+2}|   || \bm{\Delta} \mathbf{Q}^{-}_k(t)||,
	\end{aligned}
\end{equation}
where $c(z, \lambda) = \frac{2}{|r \lambda \gamma(z)|}\big( Q_0 + |r \lambda| + |z| \big)^2 $. On the other hand, we have  $c(z, \lambda) \to \infty$ when $\lambda \to 0$, or when $\gamma(z)\to 0$ at $z = \pm z_1, \, \pm z_2$ with
\begin{equation}
	\begin{aligned}
		&z_1 = \sqrt{1-2 Q^2_0 - 2 Q_0 \sqrt{Q^2_0 - 1 }}, \quad  z_2 =  \sqrt{1-2 Q^2_0 + 2 Q_0 \sqrt{Q^2_0 - 1}}.
	\end{aligned}
\end{equation}
In order to avoid $c(z) \to \infty$, we restrict our attention to the domain $(D^{-})_{\epsilon} = D^{-} \setminus (B_{\epsilon}( z_1) \cup B_{\epsilon}(- z_1) \cup B_{\epsilon}(z_2)\cup B_{\epsilon}(- z_2))$, where $D^{-} = \{\lambda \in \mathbb{C}: |\lambda|> 1 \} $, $B_{\epsilon}(\pm z_{i}) = \{z \in \mathbb{C}: |z \mp  z_{i}| < \epsilon \}$ with $\epsilon>0$ and $i = 1,2$.
It is straightforward to show that $c_{\epsilon} = \max_{z \in (D^{-})_{\epsilon}} c(z) < \infty$.

Next, by induction, we prove that for all  $z \in (D^{-})_{\epsilon}$ and for all $j \in \mathbb{N}$,
\begin{equation}
	\label{xcxa43}
	\begin{aligned}
		&|| w^{(j)}_n(z, \lambda)|| \leq \frac{\mathbf{M}^{j}_n (z, \lambda) }{j!},
	\end{aligned}
\end{equation}
where
\begin{equation}
	\begin{aligned}
		\mathbf{M}_n(z, \lambda) = \displaystyle\sum_{k = -\infty}^{n-1}  2 c_{\epsilon}    || \bm{\Delta} \mathbf{Q}^{-}_k(t)||.
	\end{aligned}
\end{equation}
The case $j=0$ is trivial.
Also, noticing that $|\lambda|>1$ for all $\lambda \in (D^{-})_{\epsilon}$, one has $ 1 +\lambda^{-2n+2k+2} \leq 2$.  Then, if \eref{xcxa43} holds for $||w^{(j)}_n(z, \lambda)||$, we have
\begin{equation}
	\label{A10}
	\begin{aligned}[b]
		||w^{(j+1)}_n (z, \lambda)|| &\leq 2 c_{\epsilon} \displaystyle\sum_{k = -\infty}^{n-1}   || \bm{\Delta} \mathbf{Q}^{-}_k(t)|| \frac{\mathbf{M}^{j}_k (z, \lambda) }{j!} \noindent \\
		& = \frac{ 2 c_{\epsilon} }{j!}  \displaystyle\sum_{k = -\infty}^{n-1}   || \bm{\Delta} \mathbf{Q}^{-}_k(t)||    \Big( \displaystyle\sum_{m = -\infty}^{k-1} 2 c_{\epsilon} || \bm{\Delta} \mathbf{Q}^{-}_m||\Big)^{j}.
	\end{aligned}
\end{equation}
Using the following inequality
\begin{equation}
	\begin{aligned}
		\displaystyle\sum_{k = -\infty}^{n-1} b_{k} \Big( \displaystyle\sum_{m = -\infty}^{k-1} b_{m} \Big)^{j}
		&\leq \frac{1}{j+1}  \displaystyle\sum_{k = -\infty}^{n-1}  b_{k}^{j + 1} ,
		\quad b_k>0,
	\end{aligned}
\end{equation}
Eq.~\eref{A10} can be written as
\begin{equation}
	\begin{aligned}
		|| w^{(j+1)}_n(z, \lambda)||
		&\leq \frac{ \mathbf{M}^{j+1}_{n}(z, \lambda)}{(j+1)!},
	\end{aligned}
\end{equation}
which proves the induction step. Thus,  if $\Delta \mathbf{Q}_{n}^{\pm}(t) \in \Big\{ f_n \big| \displaystyle\sum_{j = \mp \infty}^{n}  | f_{j} | < \infty, \forall n \in \mathbb{Z} \Big\}$, the summation in~\eref{xcxa39a} is absolutely and uniformly summable, which immediately implies that $\mu^{-}_{n,1}$  is analytic for $z \in D^{-}$. The rest of the theorem can be proved in a similar way.

\subsection{Proof of Eq.~\eref{hlj3.14}}
\label{appendixA3a}
First, Eq.~\eref{hlj2.53} includes the following constraint
\begin{equation}
	\label{hlj3.8}
	\begin{aligned}
		&s_{12}(\zeta_i)  = - \sigma \frac{Q_+^*}{Q_+ } s_{21}(\hat{\zeta}_i),
	\end{aligned}
\end{equation}
which, by virtue of Eq.~\eref{ss22}, immediately implies the first constraint in Eq.~\eref{hlj3.14}.

On the other hand, Eq.~\eref{hlj2.23} suggests the following relation at $\zeta = \zeta_i$ and $\hat{\zeta}_i$:
\begin{subequations}
	\label{hlj3.10}
	\begin{align}
		&\bm{\phi}^{-}_{-n,2}(t,\zeta_i) = s_{12}(\zeta_i) \bm{\phi}^{+}_{-n,1}(t,\zeta_i),\label{hlj3.10b} \\
		&\bm{\phi}^{-}_{-n,1}(t,\hat{\zeta}_i) = s_{21}(\hat{\zeta}_i) \bm{\phi}^{+}_{-n,2}(t, \hat{\zeta}_i). \label{hlj3.10a}
	\end{align}
\end{subequations}
By replacing $\bm{\phi}^{-}_{-n,2}(t,\zeta_i)$ in Eq.~\eref{hlj3.10b} through Eq.~\eref{hlj106a}, we have
\begin{subequations}
	\label{hlj311}
	\begin{align}
		&- \frac{c^*_{\infty} \Upsilon^*_{n}(t) }{\mi r \lambda_i} \big[\phi^{+}_{n+1,21}(t,-\zeta^*_{i})\big]^* = s_{12}(\zeta_i) \phi^{+}_{-n,11}(t,\zeta_{i}), \label{hlj3.11a}\\
		&  \frac{\sigma  c^*_{\infty} \Upsilon^*_{n}(t)}{\mi r  \lambda_i} \big[\phi^{+}_{n+1,11}(t,-\zeta^*_{i})\big]^* = s_{12}(\zeta_i) \phi^{+}_{-n,21}(t,\zeta_{i}). \label{hlj3.11b}
	\end{align}
\end{subequations}
Subsequently, changing $n \to -n-1$ and $(z^*_{i}, -\lambda^*_{i}, -\zeta^*_{i}) \to (z_{i}, \lambda_{i}, \zeta_{i})$ (which corresponds to the mapping in the second symmetry) in Eq.~\eref{hlj3.11b} and conjugating the resulting equation, we obtain
\begin{equation}
	\label{hlj312}
	\begin{aligned}
		&  \frac{\sigma  c_{\infty} \Upsilon_{-n-1}(t)}{\mi r \lambda_{i} } \phi^{+}_{-n,11}(t,\zeta_{i}) = s^*_{12}(-\zeta^*_i) \big[\phi^{+}_{n+1,21}(t,-\zeta^*_{i})\big]^*.
	\end{aligned}
\end{equation}
By comparing Eq.~\eref{hlj3.11a} with Eq.~\eref{hlj312}, one can derive
\begin{equation}
	\label{A18}
	\begin{aligned}
		&-\frac{\sigma  c_{\infty} \Upsilon_{-n-1}(t)}{\mi r \lambda_{i} }
		  \frac{c^*_{\infty} \Upsilon^*_{n}(t) }{\mi r \lambda_i}
		   = s_{12}(\zeta_i) s^*_{12}(-\zeta^*_i),
	\end{aligned}
\end{equation}
which, via the definitions of $c_{\infty}$ in Eq.~\eref{hlj2.25} and $\Upsilon_{n}(t)$ in Eq.~\eref{hlj2.3}, can be written as
\begin{equation}
	\label{hlj3.12a}
	\begin{aligned}
		s_{12}(\zeta_{i}) s^*_{12}(-\zeta^*_{i}) = - \sigma \lambda^{-2}(\zeta_{i})  c_{\infty}
		= -  \frac{\sigma (\mi r \zeta_{i} -1)c_{\infty} }{\zeta_{i}(\zeta_{i} - \mi r)} .
	\end{aligned}
\end{equation}
Repeating the above process to Eq.~\eref{hlj3.10a} derives
\begin{equation}
	\label{hlj3.13a}
	\begin{aligned}
		s_{21}(\hat{\zeta}_i) s^*_{21}(-\hat{\zeta}^*_i) = -
		\frac{ \sigma \hat{\zeta}_i (\hat{\zeta}_i - \mi r) c_{\infty}}{\mi r \hat{\zeta}_i -1} .
	\end{aligned}
\end{equation}
Again, based on Eq.~\eref{ss22},
one can find that Eqs.~\eref{hlj3.12a} and~\eref{hlj3.13a} are equivalent to the second and third constraints in Eq.~\eref{hlj3.14}.

\end{appendix}

\end{document}